\documentclass[version=preprint,floatrow]{iacrcc}
\PassOptionsToPackage{svgnames}{xcolor}

\license{CC-by}

\usepackage[svgnames]{xcolor}
\usepackage{amsmath,amssymb,mathtools} 
\usepackage{enumitem}
\usepackage{listings}
\usepackage[labelfont=bf]{caption} 
\usepackage{tabularray}
\usepackage{booktabs}

\definecolor{MyGreen}{HTML}{008000}
\definecolor{MyBlue}{HTML}{0000FF}

\lstdefinelanguage{DPUasm}{
  morekeywords={move, call, jump, jgtu,
    and, or, xor, add, lsl_add, sub, lsl, lsr,
    mul_step, mul_ul_ul, mul_uh_ul, mul_uh_uh},
  sensitive=true,
  morestring=[b]",
  morecomment=[l]//,
}
\lstdefinestyle{DPUasm}{
  language=DPUasm,
}

\definecolor{light_blue}{HTML}{77AADD}
\definecolor{orange}{HTML}{EE8866}
\definecolor{light_yellow}{HTML}{EEDD88}
\definecolor{pink}{HTML}{FFAABB}
\definecolor{light_cyan}{HTML}{99DDFF}
\definecolor{mint}{HTML}{44BB99}
\definecolor{pear}{HTML}{BBCC33}
\definecolor{olive}{HTML}{AAAA00}
\definecolor{pale_grey}{HTML}{DDDDDD}
\definecolor{light_grey}{HTML}{4f4f4f}

\usepackage{xspace}
\newcommand{\eg}{\emph{e.g.}\xspace}
\newcommand{\ie}{\emph{i.e.}\xspace}

\newcommand{\Z}{\mathbb Z} 

\newcommand{\BigO}{\mathcal{O}}

\newcommand{\floor}[1]{\left\lfloor#1\right\rfloor}

\newcommand{\sqbr}[1]{\left[#1\right]}

\title[
       running  = {DRAMatic: Accelerating HE on a PIM System},
      ]{DRAMatic Speedup: Accelerating HE Operations on a Processing-in-Memory System}

\addauthor[
           orcid   = {0009-0003-6752-5274},
           inst    = {1},
           email   = {n.klinger@uni-luebeck.de},
           surname = {Klinger},
          ]{Niklas Klinger}
\addauthor[
           orcid   = {0009-0007-9402-823X},
           inst    = {1},
           email   = {j.sander@uni-luebeck.de},
           surname = {Sander},
          ]{Jonas Sander}

\addauthor[
           orcid   = {0000-0002-3371-9228},
           inst    = {2},
           email   = {peterson.yuhala@unine.ch},
           surname = {Yuhala},
          ]{Peterson Yuhala}
\addauthor[
           orcid   = {0000-0003-1574-6721},
           inst    = {2},
           email   = {pascal.felber@unine.ch},
           surname = {Felber},
          ]{Pascal Felber}

\addauthor[
           orcid   = {0000-0003-1116-6973},
           inst    = {1},
           email   = {thomas.eisenbarth@uni-luebeck.de},
           surname = {Eisenbarth},
          ]{Thomas Eisenbarth}

\authorrunning{Niklas Klinger, Jonas Sander, Peterson Yuhala, Pascal Felber, Thomas Eisenbarth}

\addaffiliation[ror = 00t3r8h32,
                street     = {Ratzeburger Allee 160},
                city       = {L\"ubeck},
                postcode   = {23562},
                country = {Germany}
               ]{University of Luebeck}
               
\addaffiliation[ror = 00vasag41,
                street     = {Av. du 1er-Mars 26},
                city       = {Neuch\^atel},
                postcode   = {2000},
                country = {Switzerland}
               ]{University of Neuch\^atel}

\begin{document}

\maketitle



\begin{abstract}
Homomorphic encryption (HE) is a promising technology for confidential cloud computing,
as it allows computations on encrypted data.
However, HE is computationally expensive and often memory-bound on conventional computer architectures.
Processing-in-Memory (PIM) is an alternative hardware architecture that integrates processing units and memory on the same chip or memory module.
PIM enables higher memory bandwidth than conventional architectures and could thus be suitable for accelerating HE.
We present DRAMatic, which implements operations foundational to HE on UPMEM PIM -- a programmable general-purpose PIM system developed by UPMEM.
DRAMatic incorporates many arithmetic optimizations, including residue number system and number-theoretic transform techniques, and can support the large parameters required for secure homomorphic evaluations.
It achieves a 334 times speed-up compared to previous HE implementations on UPMEM PIM.
We also evaluate DRAMatic against Microsoft SEAL, a popular open-source HE library, regarding both runtime and energy efficiency.
The results show that DRAMatic significantly closes the gap between Microsoft SEAL and HE implementations on UPMEM PIM.
However, we also show that DRAMatic is currently constrained by data transfer overhead and limited multiplication performance on UPMEM PIM hardware. Finally, we discuss potential hardware extensions to UPMEM PIM.
\end{abstract}

\begin{textabstract}
Homomorphic encryption (HE) is a promising technology for confidential cloud computing,
as it allows computations on encrypted data.
However, HE is computationally expensive and often memory-bound on conventional computer architectures.
Processing-in-Memory (PIM) is an alternative hardware architecture that integrates processing units and memory on the same chip or memory module.
PIM enables higher memory bandwidth than conventional architectures and could thus be suitable for accelerating HE.
We present DRAMatic, which implements operations foundational to HE on UPMEM PIM -- a programmable general-purpose PIM system developed by UPMEM.
DRAMatic incorporates many arithmetic optimizations, including residue number system and number-theoretic transform techniques, and can support the large parameters required for secure homomorphic evaluations.
It achieves a 334 times speed-up compared to previous HE implementations on UPMEM PIM.
We also evaluate DRAMatic against Microsoft SEAL, a popular open-source HE library, regarding both runtime and energy efficiency.
The results show that DRAMatic significantly closes the gap between Microsoft SEAL and HE implementations on UPMEM PIM.
However, we also show that DRAMatic is currently constrained by data transfer overhead and limited multiplication performance on UPMEM PIM hardware. Finally, we discuss potential hardware extensions to UPMEM PIM.
\end{textabstract}

\section{Introduction}
Data security is becoming increasingly important as more applications rely on cloud computing, especially in areas such as healthcare, where confidentiality and privacy of sensitive data is critical.
Homomorphic encryption (HE) allows computations to be performed on encrypted data without decrypting it and without revealing any information about the inputs and outputs of the computation apart from their lengths.
The results can be decrypted using the secret key corresponding to the original encrypted data.
Current HE schemes rely on noise which is added to the encrypted data. When performing computations on this data, the noise compounds, which poses a limit on how many operations can be performed, as too much noise makes the results impossible to decrypt. Fully homomorphic encryption (FHE) solves this problem by introducing a special noise-reducing operation called bootstrapping. FHE allows arbitrary sequences of operations to be performed on the encrypted data, as long as bootstrapping is performed at appropriate times.

Because HE allows us to perform computations on encrypted data, it can significantly enhance security in cloud computing contexts.
Using HE, even highly sensitive computations can be outsourced to cloud providers or other third parties, because the data can stay encrypted throughout the whole process.
Compared to Trusted Execution Environments (TEEs), HE provides pure cryptographic security guarantees and does not require trusting the TEE hardware used by a cloud provider.
Additionally, the secret key stays on the client side and is thus not exposed to potential side-channel attacks in the cloud.
Past research has revealed many such side-channels vulnerabilities in TEEs \cite{CHES:MogIraEis17, CCS:WilSieEis24, tee-fail},
making this a significant advantage for HE.

However, the main drawbacks of HE are its high memory and compute requirements compared to working directly on the unencrypted data.
In traditional computer architectures, HE can also become bottlenecked by memory accesses,
because HE operations have comparatively low \textit{arithmetic intensity} \cite{EPRINT:CAYCVJ21}, meaning that they access a lot of data, but
perform few arithmetic operations on each piece of data.
Processing-in-Memory (PIM) is a hardware architecture that integrates processing units and memory on the same chip or memory module \cite{Ghose2019-PIM-Perspective}.
It enables low-latency, high-bandwidth memory access compared to traditional architectures, in which processing units and memory are separated.
PIM could thus be suitable for accelerating memory-intensive workloads like HE.
While other work has mostly explored custom PIM designs \cite{DBLP:Gupta2021-HEcustomPIMchips, Zhou2025-FHEmem-Accelerator},
we will focus on the general-purpose PIM system developed by UPMEM \cite{DBLP:GomezMutlu2021-BenchmarkingPIM}.
UPMEM PIM is programmable, highly parallel and provides high total memory bandwidth.
These properties could allow UPMEM PIM to break the memory-wall that is faced by other HE implementations, but without requiring fully custom hardware; shifting HE further towards being compute-bound instead of memory-bound.

However, UPMEM PIM also presents new challenges,
like overcoming the weak multiplication performance of the individual Data Processing Units (DPUs) and minimizing the communication costs between them.
In this work, we implement typical HE operations on
the general-purpose UPMEM PIM system and compare its performance with other implementations.

\subsection{Our Contribution}
We present DRAMatic, which implements typical HE operations on UPMEM PIM, namely BGV multiplication \cite{ITCS:BraGenVai12}, number-theoretic transforms (NTT), and inverse transforms (iNTT) \cite{Satriawan2023-Review-on-NTT}.
To implement BGV multiplication efficiently, we optimize polynomial operations on UPMEM PIM,
utilizing Residue Number Systems (RNS), NTT, Barrett reduction \cite{C:Barrett86},
custom multiplications and careful buffering to speed-up modular polynomial multiplication.

Compared to previous work on accelerating HE with UPMEM PIM \cite{DBLP:Unine-UPMEM-HE},
we achieve a 334 times speed-up and support larger parameter sizes, that allow secure homomorphic evaluations in practical use-cases.
Unlike concurrent work on accelerating NTT execution on UPMEM PIM \cite{2026-NTT-UPMEM-PIM},
DRAMatic does not rely on expensive inter-DPU communication
and scales better regarding the number of DPUs in the system.
We also utilize smaller RNS moduli, thus avoiding large integer arithmetic and enabling optimized multiplication routines.
Finally, DRAMatic is not limited to (i)NTT, but supports multiple operations, including modular addition and modular multiplication (polynomial or pointwise), NTT, and iNTT, which can be composed into larger operations like BGV multiplication.

Additionally, we compare the performance between DRAMatic and an optimized HE implementation on CPUs,
significantly closing the gap compared to previous work \cite{DBLP:Unine-UPMEM-HE}.
We also show that the data transfer overhead and weak multiplication performance of DPUs remains a bottleneck for DRAMatic and discuss potential solutions
in the form of hardware extensions to UPMEM PIM.
Finally, we perform the first power measurements of UPMEM PIM for HE operations and compare the energy efficiency of DRAMatic with an optimized CPU implementation. Our implementation of DRAMatic is open source\footnote{
\url{https://github.com/UzL-ITS/DRAMatic-Speedup}
}.

\section{Background}
We now go into more detail on UPMEM PIM and HE.

\subsection{UPMEM PIM}
\begin{figure}[t]
    \centering
    \includegraphics[width=\textwidth]{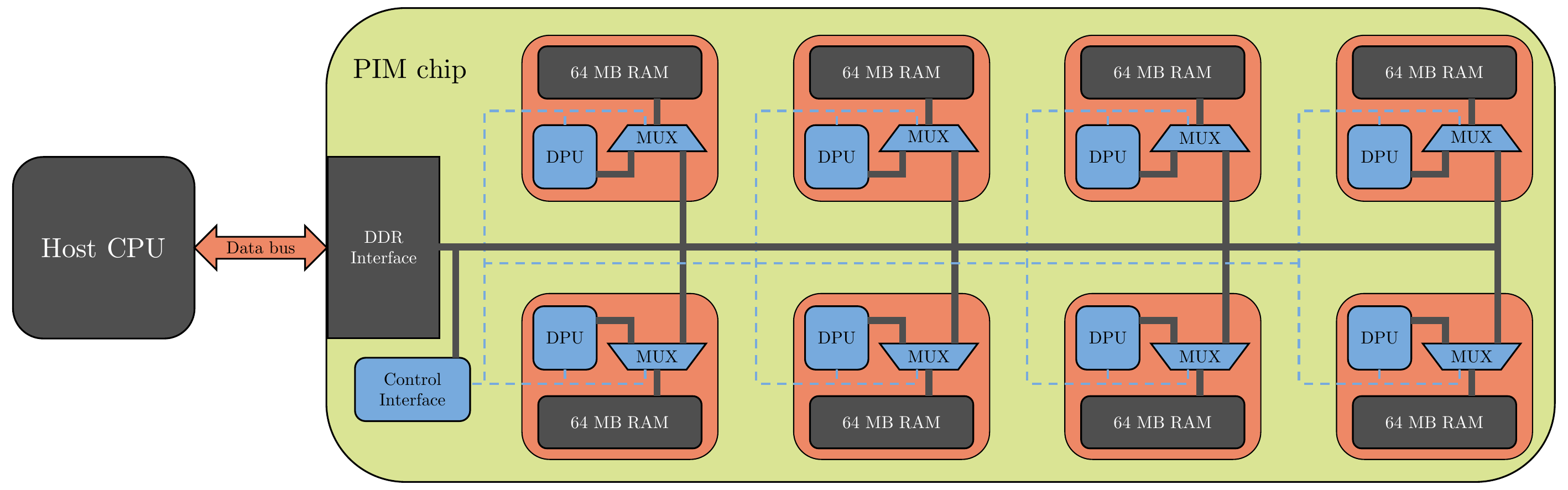}
    \caption{Architecture of an UPMEM PIM chip. Inspired by \cite{UPMEM2024-Keynote}.}
    \label{fig:diagram-PIM-chip}
\end{figure}

UPMEM PIM is a programmable near-memory computing architecture in which simple, general-purpose processors, called Data Processing Units (DPUs), are co-located with DRAM chips on special memory modules (PIM DIMMs).
The DPUs are custom 32-bit RISC processors running at up to 400 MHz.
They support 16 independent threads\footnote{These capabilities refer to the v1B DPU model. UPMEM also provides a v1A model with slight differences. See \url{https://sdk.upmem.com/2025.1.0/03_ProgrammingWithUpmemDpu.html\#dpu-chip-characteristics}.
}
and up to 11 of these threads can run concurrently at any given time.
Each PIM DIMM has a total capacity of 8 GB and contains 128 DPUs on multiple PIM chips, with each DPU having access to a 64 MB slice of the module’s main RAM (MRAM).
\autoref{fig:diagram-PIM-chip} shows the architecture of such an UPMEM PIM chip.
Current UPMEM platforms support up to 20 PIM DIMMs, for a total of 2560 DPUs and a capacity of 160 GB.

DPUs can communicate with the host CPU via the DDR interface.
Inter-DPU communication is not possible directly and must go through the host CPU instead, meaning that the data must first be copied from one DPU to the CPU’s main memory and then copied from the main memory to another DPU. This process is slow, as it requires synchronization between all three components.
Frequent inter-DPU communication should thus be avoided, which we need to consider when implementing HE operations.
The CPU can also not use the DPUs' memory directly. Instead, it must use normal DRAM for its computations and then copy the data to the PIM DIMMs when required.

Although DPUs are connected to 64 MB of MRAM, they can only directly access an additional, smaller memory, called the working memory (WRAM), of which each DPU has 64 KB. Special direct memory access (DMA) instructions are used to transfer data between the WRAM and MRAM of a DPU.
Since the memory of individual DPUs is limited, we will need to split data and inputs over multiple DPUs when implementing HE operations on UPMEM PIM.

The native word size of DPUs is 32-bit, but the DPU hardware does not directly support 32x32-bit or even 16x16-bit integer multiplications.
The hardware multiplier can only perform 8x8-bit multiplications, which yield 16-bit results.
Larger multiplications thus need to be constructed out of multiple instructions.
As implemented by the DPU compiler, 32-bit multiplications are up to 43 times slower than 8-bit multiplications; and 64-bit multiplications are 134 times slower.
This poses a challenge when implementing HE operations, which often involve big coefficients (hundreds of bits).
The DPUs also do not have hardware supported floating point operations; they are implemented in software instead, which results in poor performance. As such, floating point operations should be avoided for most applications.
This also means that optimized HE variants like \cite{RSA:HalPolSho19}
which rely on floating-point arithmetic, cannot be used on UPMEM PIM.

DPUs feature a 14-stage pipeline, but the three first and last stages of dependent instructions can execute in parallel \cite{DBLP:GomezMutlu2021-BenchmarkingPIM}.
Thus, only 11 active threads are needed to fully saturate the pipeline.
Active threads are scheduled in a round-robin way; with
threads considered active unless they are explicitly waiting (\eg, on a mutex) or performing a DMA operation.
These DMA operations execute sequentially; and threads waiting on a DMA operation are placed in a special queue. Thus, only one thread of each DPU can access the MRAM at any given time.

\subsection{HE Schemes}
Current practical HE schemes are based on the Learning With Errors (LWE) or Ring Learning With Errors (RLWE) problem and use noisy polynomials as ciphertexts.
RLWE based HE schemes like
BGV \cite{ITCS:BraGenVai12} operate on polynomial rings denoted as $\Z_q[x]/(x^n+1)$, \ie, the ring of integer polynomials modulo the polynomial $x^n+1$ and with coefficients modulo $q$
(other rings are possible, but this is the most popular choice).
Ciphertexts then consist of two or more such polynomials.
The two main operations on ciphertexts, which make the schemes homomorphic, require polynomial addition and polynomial multiplication. It may also be required to perform polynomial modulus and coefficient modulus operations, to ensure that the ciphertexts stay within the relevant polynomial ring.

\subsubsection{Steps in an HE Scheme}
\begin{figure}[t]
    \centering
    \includegraphics[width=\textwidth]{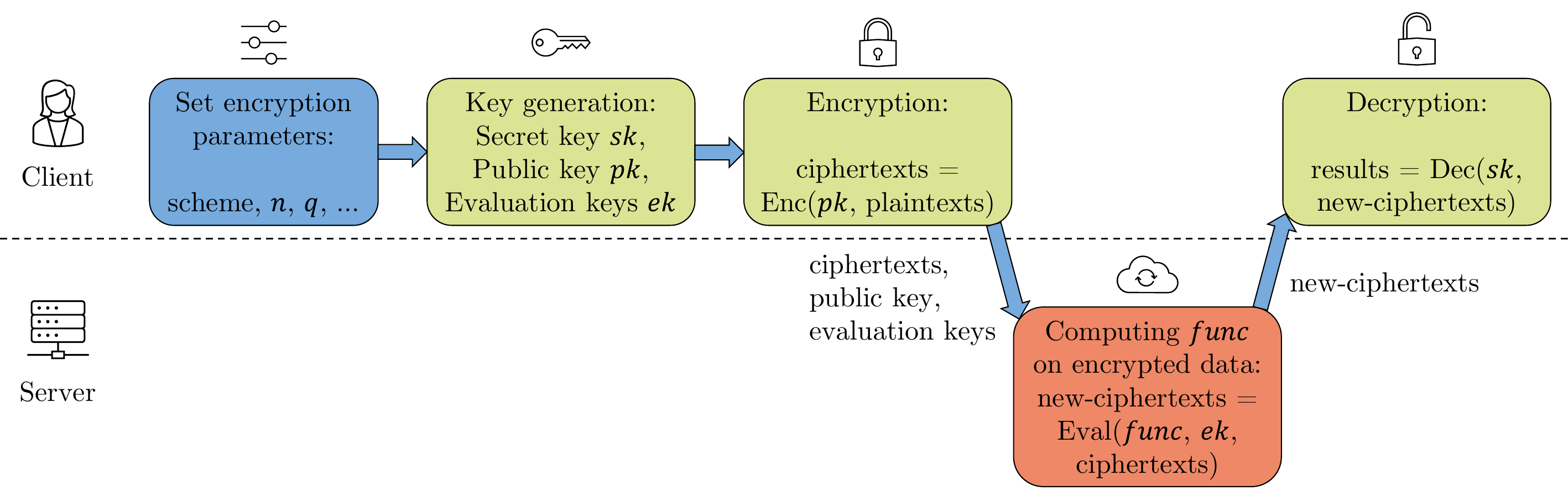}
    \caption{Steps in the usage of an HE scheme, showing the roles of client and server.}
    \label{fig:HE-steps}
\end{figure}

Usage of an HE scheme can be logically split into the following steps: parameter selection, key generation, encryption, computation, and decryption, as shown in \autoref{fig:HE-steps}.
In this work, we consider a scenario in which a client wants to offload computations to an untrusted server (\eg, a cloud service provider) using HE.
This means that key generation, encryption, and decryption are performed by the client,
while the server handles computations on the encrypted data and potentially negotiates encryption parameters with the client.
Our focus is the computation on encrypted data, corresponding to the server side of this interaction.

\subsubsection{BGV}
For our evaluation, we focus on the BGV multiplication.
In BGV, two ciphertexts $ct = (ct_0, ct_1)$ and $ct' = (ct'_0, ct'_1)$, are multiplied to produce $ct_{res} = (c_0, c_1, c_2)$ as follows:
\begin{align*}
    c_0 &= \sqbr{ct_0 \cdot ct'_0}_{q_l} \\
    c_1 &= \sqbr{ct_0 \cdot ct'_1 + ct_1 \cdot ct'_0}_{q_l} \\
    c_2 &= \sqbr{ct_1 \cdot ct'_1}_{q_l} \\
\end{align*}
where $[\cdot]q_l$ is a modular reduction by the current modulus level, which can
change through a process called \emph{modulus switching}.
In BGV, modulus switching improves the performance of the scheme and is done when ciphertexts become too noisy after a series of operations.

Note that the BGV multiplication does not involve scaling or rounding operations as in other schemes like BFV.
When we combine BGV with RNS, this means that all operations required for BGV multiplication can be performed on the individual residues independently from each other.
BGV multiplication is thus well suited for UPMEM PIM,
as it allows us to split a multiplication over multiple DPUs using RNS, which we describe below.

\section{Design Considerations}
In the following, we discuss the challenges of efficiently implementing BGV operations on UPMEM PIM DPUs.

\subsection{Large Coefficients}
To efficiently operate on the large coefficients of HE ciphertexts, we use a RNS representation for the ciphertext polynomials.
An RNS enables the decomposition of large integers into several smaller residues, each computed modulo a distinct base.
In this representation, a large modulus $M$ is expressed as the product of smaller pairwise coprime moduli $m_1, m_2, \ldots, m_k$, called the RNS base.
Each coefficient $x$ modulo $M$ is then represented by the tuple of its residues
\[
(x_1, x_2, \ldots, x_k), \quad \text{where } x_i = x \bmod m_i.
\]
Addition and multiplication can now be performed independently on each residue, \ie,
\[
(x \circ y) \bmod M \;\longleftrightarrow\; (x_1 \circ y_1 \bmod m_1,\, x_2 \circ y_2 \bmod m_2,\, \ldots,\, x_k \circ y_k \bmod m_k),
\]
where $\circ$ denotes either addition or multiplication.
The original value modulo $M$ can be reconstructed from its residues using the Chinese Remainder Theorem (CRT):
\[
x \equiv \sum_{i=1}^k x_i M_i N_i \pmod{M},
\]
where $M_i = M / m_i$ and $N_i$ is the modular inverse of $M_i$ modulo $m_i$.
For the mathematical background, we refer the reader to works on number theory and algebra, \eg, \cite{Shoup2005-NumberTheoryAlgebra}.

Crucially, RNS enables us to do computations modulo $M$, by operating on smaller numbers moduli $m_i$.
Because DPU multiplication and modulus performance is limited, we want the individual residues to be as small as possible.
For our use-case, this results in 32-bit residues, which is also the DPUs' native word size.
16-bit residues would be even faster, but they cannot be combined with NTT, as there aren't enough suitable primes smaller than 16-bit (see \autoref{sec:combining-RNS-NTT}).
Additionally, we implement a custom 32-bit multiplication routine, which is faster than the compiler generated one for our use-case and reduces the speed difference between 32-bit multiplication and 16-bit multiplication.

\subsubsection{Custom Multiplication Routines}
We improve multiplication performance by implementing custom 32-bit multiplication routines on the DPUs.
While we can limit our inputs to 32-bit using RNS, we still need 64-bit intermediate results to correctly perform modular reductions.
When using native arithmetic as implemented by the DPU compiler, we have to perform full 64-bit multiplications to get correct results, even though our inputs are only 32-bit values.
Our custom multiplication routine instead constructs the result by appropriately shifting and adding the results of four native 16x16-bit multiplications,
as shown in \autoref{lst:custom_mul_32to64}.
This straightforward routine takes 35 cycles and is about 70\% faster than the native full 64-bit multiplication.
We also tested a Karatsuba implementation to reduce the number of 16x16-bit multiplications to three, but this is actually 4 cycles slower, as the additional overhead for combining the results is more significant than the saved multiplication.

\begin{figure}[t]
\RawFloats
\centering
\begin{minipage}{0.45\textwidth}%
    \centering%
    \footnotesize%
    \begin{lstlisting}[label={lst:custom_mul_32to64}, captionpos=b, caption={Pseudocode of our custom 32-bit to 64-bit multiplication routine. Combining the results is done in hand-written assembly for increased performance.}]
fn custom_mul_32to64(a, b):
  a0, a1 = split(a, 16)
  b0, b1 = split(b, 16)
  lo = a0 * b0
  m1 = a0 * b1
  m2 = a1 * b0
  hi = a1 * b1
  // combine results
  return hi << 32 +
    (m1 + m2) << 16 + lo
    \end{lstlisting}%
\end{minipage}\hfill\begin{minipage}{0.5\textwidth}%
    \centering%
    \includegraphics[]{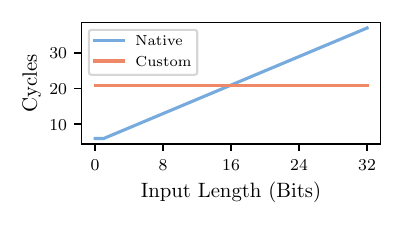}%
    \vspace{-9pt}
    \captionof{figure}{Runtime of our custom 32-bit to 32-bit multiplication and the native DPU implementation for different input sizes.}%
    \label{fig:plot_custom_mul}%
\end{minipage}
\end{figure}

Similarly, we also implement custom 32-bit multiplication that only produces 32-bit results.
In that routine, we skip the \texttt{hi}-computation from \autoref{lst:custom_mul_32to64} and simplify the computations of \texttt{m1} and \texttt{m2}, as their upper bits are unused, which reduces the runtime to 21 cycles.
Note that the runtime of the native 32-bit multiplication with 32-bit results depends on the highest set bit in its inputs. The native implementation makes a compromise: it is optimal for small inputs, which are typical in many applications, but is relatively slow in the worst case.
Our custom multiplication routine has constant runtime instead.
The result is shown in \autoref{fig:plot_custom_mul}; our implementation is faster when both factors are longer than 16 bits, but is otherwise slower than the native implementation.
However, this is still an improvement for our use-case;
since we are operating on encrypted data, large input values are common, as the data is seemingly random.

\subsubsection{Barrett Reduction}
To further speed up modular multiplication, we utilize Barrett reduction, which is optimized for repeated reductions by the same modulus
and was first introduced by Barrett \cite{C:Barrett86} for a fast RSA implementation.
Given $v$ and $m$, consider that we want to find the remainder $x = v \bmod m$.
The idea of Barrett reduction is to pre-compute a scaled reciprocal $R=\floor{2^n / m}$ of the modulus.
The remainder can then be approximated as $x \approx v - m \cdot \floor{v R / 2^n}$ and can be corrected with a final conditional subtraction.
In essence, we exchange the original division for two multiplications, a bit shift and a conditional subtraction, which is much faster on DPUs.
In DRAMatic, the \textit{Barrett factor} $R$ of every modulus is pre-computed by the CPU and then transferred to the appropriate DPUs.

\subsection{Large Polynomials}
\begin{figure}[t]
    \centering
    \begin{tikzpicture}[thick, scale=0.7]
        \draw[dashed] (3.5, -0.2) -- (3.5, -8.5);
        \draw[dashed] (6.5, -0.2) -- (6.5, -8.5);
        \node at (1, -0.4) {Stage 1};
        \node at (5, -0.4) {Stage 2};
        \node at (8, -0.4) {Stage 3};
        \foreach \i [evaluate=\j using int(\i - 1)] in {1, ..., 8} {
            \coordinate (A\i) at (-2, -\i);
            \coordinate (B\i) at (-1, -\i);
            \coordinate (C\i) at (3, -\i);
            \coordinate (D\i) at (4, -\i);
            \coordinate (E\i) at (6, -\i);
            \coordinate (F\i) at (7, -\i);
            \coordinate (G\i) at (9, -\i);
            \coordinate (H\i) at (10, -\i);

            \draw (A\i) -- (B\i);
            \draw (C\i) -- (D\i);
            \draw (E\i) -- (F\i);
            \draw[->] (G\i) -- (H\i);

            \node[anchor=east] at (A\i) {$c_{\j}$};
        }
        \node[anchor=west] at (H1) {$t_0$};
        \node[anchor=west] at (H2) {$t_4$};
        \node[anchor=west] at (H3) {$t_2$};
        \node[anchor=west] at (H4) {$t_6$};
        \node[anchor=west] at (H5) {$t_1$};
        \node[anchor=west] at (H6) {$t_5$};
        \node[anchor=west] at (H7) {$t_3$};
        \node[anchor=west] at (H8) {$t_7$};
        
        \foreach \i [evaluate=\j using \i + 4] in {1, ..., 4} {
            \draw (B\i) -- (C\j);
            \draw (B\j) -- (C\i);
            \node[anchor=south] at (C\i) {$+$};
            \node[anchor=south] at (C\j) {$-$};
            \node[anchor=south] at (B\j) {$\psi^4$};
        }
        
        \foreach \i [evaluate=\j using \i + 2] in {1, 2} {
            \draw (D\i) -- (E\j);
            \draw (D\j) -- (E\i);
            \node[anchor=south] at (E\i) {$+$};
            \node[anchor=south] at (E\j) {$-$};
            \node[anchor=south] at (D\j) {$\psi^2$};
        }
        \foreach \i [evaluate=\j using \i + 2] in {5, 6} {
            \draw (D\i) -- (E\j);
            \draw (D\j) -- (E\i);
            \node[anchor=south] at (E\i) {$+$};
            \node[anchor=south] at (E\j) {$-$};
            \node[anchor=south] at (D\j) {$\psi^6$};
        }
        
        \foreach \i [evaluate=\j using \i + 1] in {1, 3, 5, 7} {
            \draw (F\i) -- (G\j);
            \draw (F\j) -- (G\i);
            \node[anchor=south] at (G\i) {$+$};
            \node[anchor=south] at (G\j) {$-$};
        }
        \node[anchor=south] at (F2) {$\psi^1$};
        \node[anchor=south] at (F4) {$\psi^5$};
        \node[anchor=south] at (F6) {$\psi^3$};
        \node[anchor=south] at (F8) {$\psi^7$};
    \end{tikzpicture}
    
    \caption{Visualization of NTT butterflies. Shown for $n=8$ with $\log_2(n)=3$ stages.}
    \label{fig:butterfly-overview-n8}
\end{figure}
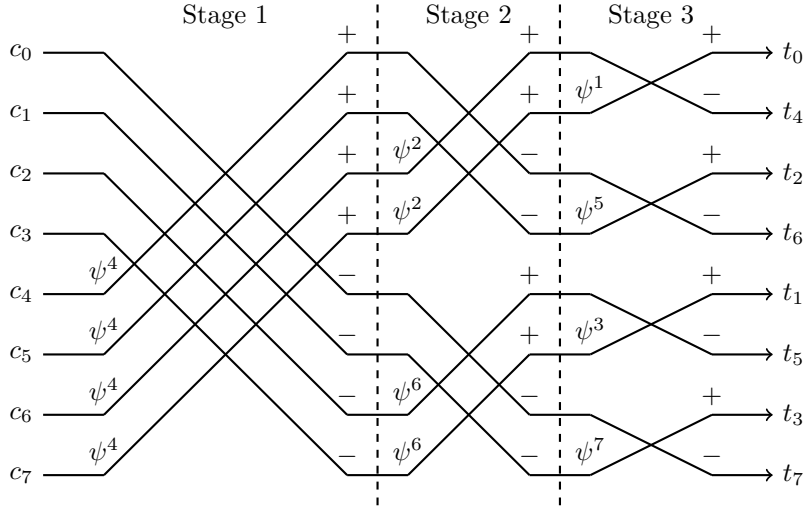

The usage of RNS representations, Barrett reduction, and our custom multiplication routine already speeds up the multiplication of individual coefficients, but the large polynomial multiplications used in HE must also be optimized.
Recall that a naive polynomial multiplication requires $\BigO(n^2)$ steps.
To improve polynomial multiplication performance,
we utilize negacyclic NTT \cite{Satriawan2023-Review-on-NTT},
which is a generalization of the discrete Fourier transform (DFT) to a finite field.
After performing NTT, polynomial multiplications (and additions) can be performed pointwise, \ie, in linear time.
An inverse NTT (iNTT) can then transform the resulting polynomial back into coefficient form.
More formally, given two polynomials $A$ and $B$ from the polynomial ring $\Z_{q}[x]/(x^n+1)$, it holds that
\[ A \cdot B = \text{iNTT}(\text{NTT}(A) \circ \text{NTT}(A)) \]
where $\cdot$ is modular polynomial multiplication and $\circ$ is modular pointwise multiplication.
Since both NTT and iNTT, can be applied in $\BigO(n\log{n})$ time, the total time required for polynomial multiplication using NTT is reduced to $\BigO(n\log{n})$.

As NTT is a generalization of the DFT, many fast DFT algorithms can also be adapted to it,
like the Cooley-Tukey fast Fourier transform \cite{CooleyTukey1965-FFT} and many of its variations.
To speed-up polynomial multiplication on DPUs,
we implement fast NTT and iNTT with bit-reversed outputs, based on Cooley-Tukey and Gentleman-Sande \cite{GentlemanSande1966-FFT} butterflies.
The idea behind these algorithms is a divide and conquer approach,
as the NTT of a length $n$ polynomial can be computed from the NTTs of its two parts with length $n/2$.
In practice, the algorithms require $\log_2(n)$ passes over the data (stages),
in which so-called butterflies are applied to pairs of coefficients.
A visualization of the NTT butterfly stages is shown in \autoref{fig:butterfly-overview-n8}.
Because of the butterflies' crossing pattern, the output of the final NTT or iNTT stage is produced in bit-reversed order. For example, on the right of \autoref{fig:butterfly-overview-n8}, output $t_4=t_{100_2}$ is actually in the $\mathsf{reverse}(100)_2=001_2=1$st position (counting from 0 at the top).
This bit-reversed order can be corrected by an additional reordering-pass.
However, BGV multiplication only requires polynomial addition and polynomial multiplication, which are pointwise operations after an NTT. Thus, the order of elements is irrelevant for these operations (as long as it is consistent) and we skip the additional reordering pass as an additional optimization.
Computations are thus performed in bit-reversed order and the results are then reversed again by the final iNTT.
Lastly, these NTTs take so-called twiddle factors as additional inputs, which are pre-computed powers of a root-of-unity $\psi$, as seen in \autoref{fig:butterfly-overview-n8}.
In DRAMatic, these twiddle factors are pre-computed by the CPU and then transferred to the DPUs.

\subsubsection{Combining RNS and NTT} \label{sec:combining-RNS-NTT}
\begin{figure}[t]
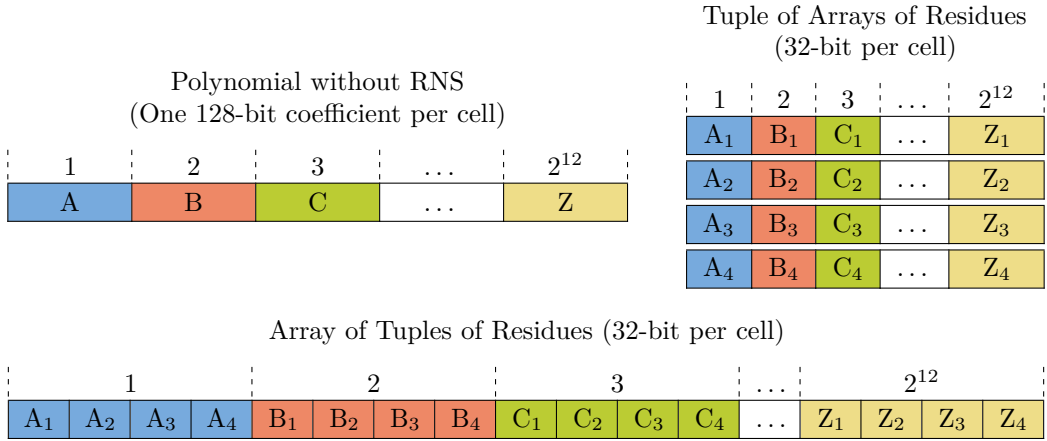

    \centering
    \begin{tblr}{
        colspec = {c c c c c},
        vline{1-6} = {2}{dashed},
        vline{1-6} = {3}{solid},
        stretch = 0,
        rows = {1em},
        columns = {3.4em},
        cell{3}{1} = {light_blue},
        cell{3}{2} = {orange},
        cell{3}{3} = {pear},
        cell{3}{5} = {light_yellow},
    }
        \SetCell[c=5]{c}{Polynomial without RNS\\(One 128-bit coefficient per cell)} \\[1ex]
        1 & 2 & 3 & \ldots & $2^{12}$ \\
        \hline
        A & B & C & \ldots & Z \\
        \hline
    \end{tblr}\hfill%
    \begin{tblr}{
        colspec = {c c c c c},
        vline{1-6} = {2}{dashed},
        vline{1-6} = {3}{solid},
        vline{1-6} = {5}{solid},
        vline{1-6} = {7}{solid},
        vline{1-6} = {9}{solid},
        stretch = 0,
        row{2, 3, 5, 7, 9} = {1em},
        cell{3, 5, 7, 9}{1} = {light_blue},
        cell{3, 5, 7, 9}{2} = {orange},
        cell{3, 5, 7, 9}{3} = {pear},
        cell{3, 5, 7, 9}{5} = {light_yellow},
    }
        \SetCell[c=5]{c}{Tuple of Arrays of Residues\\(32-bit per cell)} \\[1ex]
        1 & 2 & 3 & \ldots & $2^{12}$ \\
        \hline
        A$_1$ & B$_1$ & C$_1$ & \ldots & Z$_1$ \\
        \hline
        \\[-99pt]
        \hline
        A$_2$ & B$_2$ & C$_2$ & \ldots & Z$_2$ \\
        \hline
        \\[-99pt]
        \hline
        A$_3$ & B$_3$ & C$_3$ & \ldots & Z$_3$ \\
        \hline
        \\[-99pt]
        \hline
        A$_4$ & B$_4$ & C$_4$ & \ldots & Z$_4$ \\
        \hline
    \end{tblr}
    \\[2ex]
    \begin{tblr}{
        colspec = {c c c c c c c c c c c c c c c c c},
        vline{1-18} = {2}{dashed},
        vline{1-18} = {3}{solid},
        stretch = 0,
        rows = {1em},
        columns = {1.04em},
        cell{3}{1-4} = {light_blue},
        cell{3}{5-8} = {orange},
        cell{3}{9-12} = {pear},
        cell{3}{14-17} = {light_yellow},
    }
        \SetCell[c=17]{c}{Array of Tuples of Residues (32-bit per cell)} \\[1ex]
        \SetCell[c=4]{c} 1 & & & & \SetCell[c=4]{c} 2 & & & & \SetCell[c=4]{c} 3 & & & & \ldots & \SetCell[c=4]{c} $2^{12}$ \\
        \hline
        A$_1$ & A$_2$ & A$_3$ & A$_4$ & B$_1$ & B$_2$ & B$_3$ & B$_4$ & C$_1$ & C$_2$ & C$_3$ & C$_4$ & \ldots & Z$_1$ & Z$_2$ & Z$_3$ & Z$_4$ \\
        \hline
    \end{tblr}
    \caption{Illustration of different memory layouts for ciphertext polynomials. Shown here for $n=4096$ and 109-bit coefficients (using 128-bit or $4\times32$-bit data types). Residues with the same subscript share the same modulus and are thus assigned to the same DPU (see \autoref{sec:splitting-over-DPUs}). DRAMatic uses the tuple of arrays of residues layout (top right).}
    \label{fig:tuple-arrays-residues}
\end{figure}

To combine RNS and NTT, we split the coefficients of the ciphertext polynomials using RNS and then perform NTT on the groups of residues which share a modulus, which we also call sub-polynomials.
The polynomials are stored as multiple sub-polynomials and each sub-polynomial is continuous in memory. In other words, the polynomials are stored in a \textit{tuple of arrays of residues} layout. Compared to an \textit{array of tuples of residues} layout, this provides more flexibility in distributing the data between DPUs or DPU threads and simplifies the iteration over residues of the same modulus, because they are continuous in memory.
See \autoref{fig:tuple-arrays-residues} for an illustration.

Note that NTT requires a prime working modulus $p$, which satisfies $p = 2n + 1$ (where $n$ is the length of the polynomials). This sets a minimum size on the RNS residues, as for typical HE polynomial lengths, \eg, $n=4096$, there aren't enough such primes smaller than 16-bit. We thus use 32-bit residues in DRAMatic.

\subsection{Splitting Work over DPUs} \label{sec:splitting-over-DPUs}
We split the coefficient data over groups of DPUs based on its residues, which allows us to parallelize work (even for single ciphertext pairs) while avoiding inter-DPU communication.
Each DPU gets configured for a particular modulus of the RNS base and then only operates
on sub-polynomials of that modulus.
For example, consider an RNS base $(m_1, m_2, m_3)$.
Each polynomial of a ciphertext $ct$ would thus consist of three sub-polynomials:
\[ ct = (ct_0, ct_1) = ((ct_0^{m_1}, ct_0^{m_2}, ct_0^{m_3}), (ct_1^{m_1}, ct_1^{m_2}, ct_1^{m_3})) \]
One third of the available DPUs would then be configured for modulus $m_1$ and operate only on sub-polynomials $ct_0^{m_1}$ and $ct_1^{m_1}$ of a particular ciphertext.
Another third would be configured for modulus $m_2$ and thus operate on $ct_0^{m_2}$ and $ct_1^{m_2}$, while the last third would be configured for modulus $m_3$.
Another advantage of this split is that each DPU only has to operate on a single modulus and thus only needs support data (like NTT twiddle factors) for this one modulus, which reduces the memory requirements for the DPUs.

Note that DPU groups can only be allocated in whole ranks, which can limit the total hardware utilization for some coefficient sizes.
For example, with 8 available ranks and 3 RNS moduli, we can use 3 DPU groups of 1 rank each (3 ranks total), or 3 DPU groups of 2 ranks each (6 ranks total). However, DRAMatic cannot normally use all 8 ranks in this case.
For this reason, we also test an alternative implementation of DRAMatic, in which moduli are processed sequentially using all DPUs. For example, all 8 ranks would be configured for modulus $m_1$, then reconfigured for $m_2$, and finally for $m_3$.
This sequential implementation is slightly less efficient than the normal (parallel) implementation of DRAMatic, but it is more flexible and provides additional data points for assessing how DRAMatic's performance scales with parallelism.

\subsection{Other Considerations}
To optimize the memory layout of iNTT twiddle factors, we store them in a different, \textit{scrambled} order, similar to Microsoft SEAL \cite{sealcrypto}.
To be precise, the $i$-th inverse root of unity power $\psi^{-i}$ is stored in position $1 + \mathsf{bit\_reverse}(i-1)$, which closely matches the access patterns of our iNTT implementation on DPUs.
This order ensures that both the polynomial data and the twiddle factors are always accessed sequentially in memory, which allows for more effective buffering in DPU WRAM and thus improves performance.
We measured a 1.27 times speed-up using this scrambled order compared to the logical order.

We differentiate between coarse-grained and fine-grained multithreading on DPUs.
In fine-grained multi-threading, multiple DPU threads operate on the same sub-polynomials simultaneously. This can improve performance even with few sub-polynomials.
In coarse-grained multithreading, each DPU thread operates on its own sub-polynomials independently from other threads. For the maximum performance improvement, this type of multi-threading requires the DPU to operate on many sub-polynomials concurrently.
We use fine-grained multi-threading for element-wise operations, like modular addition or multiplication,
as these operations are easily parallelizable and require little synchronization between threads.
For more complex operations, like NTT and iNTT, we use coarse-grained multi-threading instead, which avoids synchronization overhead.
This can limit performance when operating on only a few sub-polynomials. However, in a typical HE workload, all incoming data and all results are transformed in bulk via NTT and iNTT respectively, which means that NTT operations with few sub-polynomials should be rare.

\begin{figure}[t]
    \centering
    \includegraphics[scale=0.9]{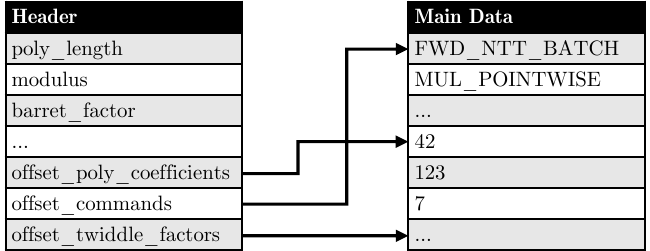}
    \vspace{0.8em}
    \caption{Visualization of the CPU-DPU interface memory layout. The main data immediately follows the header and spans the rest of the DPU’s MRAM.}
    \label{fig:InterfaceLayout}
\end{figure}

For the CPU-DPU interface, we use a single MRAM symbol.
The memory at this symbol begins with a header, followed by the \textit{main data} --- an array of DPU words that spans the rest of the DPU’s MRAM.
The header contains information about the rest of the data, like the polynomial length, the modulus, and the number of commands.
It also contains some pre-computed data, like the modular inverse of the polynomial length,
which is needed to scale down the results of an iNTT.
Lastly, the header contains offsets into the \textit{main data}, which define where different data sections start, like the twiddle factors, the commands, or the sub-polynomials to operate on. See \autoref{fig:InterfaceLayout} for a visualization.
This design is very flexible because it allows the host to dynamically define the positions and sizes of the various data sections. It also allows the host to update multiple of these sections in a single data transfer.

Due to this interface design, DRAMatic has low memory overhead on DPUs, only about the size of two sub-polynomials.
It consists of the twiddle factors and (to a lesser degree) the interface header.
DRAMatic can fill the rest of each DPU's 60 MB memory\footnote{DPUs have 64 MB of MRAM, but the UPMEM runtime requires some memory for itself, as well as for logging functions like \texttt{printf}. We reserve 4 MB for this use (about 6\%).} with sub-polynomials.
For a polynomial length of 4096, this means that DRAMatic can store about 
3750 sub-polynomials per DPU, corresponding to 1875 ciphertexts.
For a length of 8192, these numbers are halved, \ie, only 1875 sub-polynomials (937 ciphertexts), etc.

\subsection{Batching Constraints}
DRAMatic can only utilize all available DPUs when processing ciphertexts batches with a minimum size.
In the worst case, consider an UPMEM PIM system with the maximum number of 2560 DPUs and 16 threads per DPU. For coarse-grained multi-threaded operations like NTT, 
DRAMatic then needs at least $2560 \cdot 16 = 40960$ sub-polynomials per batch to fully utilize all DPUs.
The exact number of ciphertexts and memory requirements of such a batch depend on the polynomial degree and the number of RNS moduli of the corresponding coefficient size.
For example, with a polynomial degree of 4096 and 3 RNS moduli, DRAMatic would need at least $40960 \div 2 \div 3 = 6826$ ciphertexts per batch, as there are 2 polynomials per ciphertext and 3 sub-polynomials per polynomial. These ciphertexts would require 640 MB of memory.
\autoref{tab:batching-constraints} shows the batch-size requirements for different numbers of DPUs and polynomial degrees.
Note that for fine-grained multi-threaded operations like BGV multiplication, these requirements are reduced by a factor of 16.

The above requirements are lower limits for full DPU utilization, but larger batching sizes can still be useful to reduce the frequency of data transfers between DPUs and the CPU.
In a realistic scheduling scenario, multiple batches might also need to be stored simultaneously.
For example, because the host receives new ciphertexts during DPU execution and must buffer them in a new batch until the current batch completes.
In this scenario, the host CPU would require memory that is at least twice the size of the batches described above.

\begin{table}
\centering
{\footnotesize
\begin{tabular}{@{}rrrrrrr@{}}
\toprule
\multicolumn{2}{r}{\textbf{DPUs}} & 2560 & 2048 & 1536 & 1024 & 512 \\
\addlinespace[6pt]
\multicolumn{2}{r}{sub-polys. required} & 40960 & 32768 & 24576 & 16384 & 8192 \\ \midrule
\textbf{Degree} & Moduli &  &  &  &  &  \\ \midrule
1024 & 1 & \begin{tabular}[c]{@{}r@{}}20480 CTs\\ 160 MB\end{tabular} & \begin{tabular}[c]{@{}r@{}}16384 CTs\\ 128 MB\end{tabular} & \begin{tabular}[c]{@{}r@{}}12288 CTs\\ 96 MB\end{tabular} & \begin{tabular}[c]{@{}r@{}}8192 CTs\\ 64 MB\end{tabular} & \begin{tabular}[c]{@{}r@{}}4096 CTs\\ 32 MB\end{tabular} \\ \midrule
2048 & 2 & \begin{tabular}[c]{@{}r@{}}10240 CTs\\ 320 MB\end{tabular} & \begin{tabular}[c]{@{}r@{}}8192 CTs\\ 256 MB\end{tabular} & \begin{tabular}[c]{@{}r@{}}6144 CTs\\ 192 MB\end{tabular} & \begin{tabular}[c]{@{}r@{}}4096 CTs\\ 128 MB\end{tabular} & \begin{tabular}[c]{@{}r@{}}2048 CTs\\ 64 MB\end{tabular} \\ \midrule
4096 & 3 & \begin{tabular}[c]{@{}r@{}}6826 CTs\\ 640 MB\end{tabular} & \begin{tabular}[c]{@{}r@{}}5461 CTs\\ 512 MB\end{tabular} & \begin{tabular}[c]{@{}r@{}}4096 CTs\\ 384 MB\end{tabular} & \begin{tabular}[c]{@{}r@{}}2730 CTs\\ 256 MB\end{tabular} & \begin{tabular}[c]{@{}r@{}}1365 CTs\\ 128 MB\end{tabular} \\ \midrule
8192 & 6 & \begin{tabular}[c]{@{}r@{}}3413 CTs\\ 1.25 GB\end{tabular} & \begin{tabular}[c]{@{}r@{}}2730 CTs\\ 1.00 GB\end{tabular} & \begin{tabular}[c]{@{}r@{}}2048 CTs\\ 768 MB\end{tabular} & \begin{tabular}[c]{@{}r@{}}1365 CTs\\ 512 MB\end{tabular} & \begin{tabular}[c]{@{}r@{}}682 CTs\\ 256 MB\end{tabular} \\ \midrule
16384 & 13 & \begin{tabular}[c]{@{}r@{}}1575 CTs\\ 2.50 GB\end{tabular} & \begin{tabular}[c]{@{}r@{}}1260 CTs\\ 2.00 GB\end{tabular} & \begin{tabular}[c]{@{}r@{}}945 CTs\\ 1.50 GB\end{tabular} & \begin{tabular}[c]{@{}r@{}}630 CTs\\ 1.00 GB\end{tabular} & \begin{tabular}[c]{@{}r@{}}315 CTs\\ 512 MB\end{tabular} \\ \midrule
32768 & 28 & \begin{tabular}[c]{@{}r@{}}731 CTs\\ 5.00 GB\end{tabular} & \begin{tabular}[c]{@{}r@{}}585 CTs\\ 4.00 GB\end{tabular} & \begin{tabular}[c]{@{}r@{}}438 CTs\\ 3.00 GB\end{tabular} & \begin{tabular}[c]{@{}r@{}}292 CTs\\ 2.00 GB\end{tabular} & \begin{tabular}[c]{@{}r@{}}146 CTs\\ 1.00 GB\end{tabular} \\ \bottomrule
\end{tabular}
}
\caption{Batch-size requirements in number of ciphertexts (CTs) and memory for DRAMatic to fully utilize different numbers of DPUs in coarse-grained multi-threaded operations like NTT. For fine-grained multi-threaded operations, the requirements are 16 times lower.}
\label{tab:batching-constraints}
\end{table}

\section{Evaluation} \label{sec:evaluation}
We first evaluate DRAMatic's performance compared to previous work,
which also implements NTT operations on UPMEM PIM \cite{DBLP:Unine-UPMEM-HE},
which we will refer to as $\textit{MHY}^\textit{+}$.
Both implementations run on our PIM test machine, which is equipped with two Intel\textsuperscript{\textregistered} Xeon\textsuperscript{\textregistered} Silver 4216 CPUs, 256 GB of standard DDR4 memory, and 4 UPMEM PIM DIMMs with a total capacity of 32 GB, 8 DPU ranks, and 512 DPUs (509 usable).
When measuring execution time, we separately
measure the time required for data transfer, DPU computation, and data retrieval. Additionally, we also measure the power consumption of the machine using a GUDE Expert Power Control 8221 metered Power Distribution Unit (PDU).
This PDU measures the power consumption ``at the wall'', \ie, full system power, including power conversion losses.
A significant difference between $\text{MHY}^+$ and DRAMatic, is that $\text{MHY}^+$ offloads individual NTT stages onto DPUs, which has large copying and synchronization overhead,
while DRAMatic instead offloads the whole NTT operation at once.
DRAMatic also utilizes the techniques described above, like RNS, custom multiplications and careful MRAM buffering, to improve performance.
We test both implementations with 128 DPUs and a polynomial length of 2048,
with a corresponding coefficient size of 54-bit\footnote{This combination achieves 128-bit security according to the Homomorphic Encryption Standard \cite{HomomorphicEncryptionSecurityStandard}}.
We do not test longer polynomials, because $\text{MHY}^+$ only supports NTT on coefficients up to 64-bit,
which also limits the operations that can be homomorphically evaluated (see \autoref{sec:related-work-unine}).
DRAMatic does not have this limitation, but we use the same parameters here for comparability.
The raw results are listed in \autoref{sec:appendix-raw-results} (\autoref{tab:raw-results-7}).

\begin{figure}[t]
    \centering
    \includegraphics[width=0.94\linewidth]{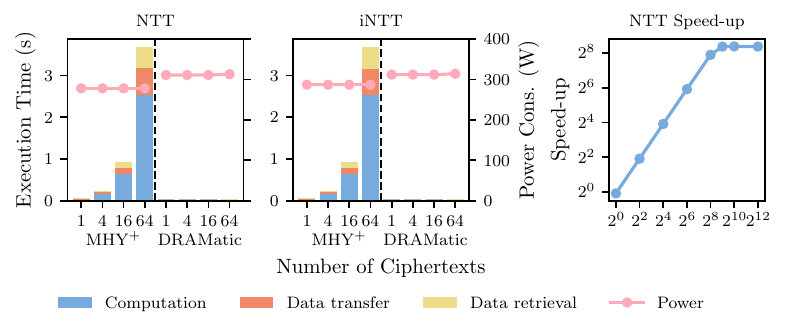}
    \caption{NTT and iNTT performance of DRAMatic compared to previous work \cite{DBLP:Unine-UPMEM-HE}. Shown for different numbers of ciphertexts. The log-log plot (right) shows how the speed-up of DRAMatic compared to $\text{MHY}^+$ increases until $2^9$ ciphertexts.}
    \label{fig:plot-vs-unine-multiple}
\end{figure}

\autoref{fig:plot-vs-unine-multiple} compares $\text{MHY}^+$ and DRAMatic for NTT and iNTT on $n$ ciphertexts ($n$ ranging from 1 to 4096).
The ciphertexts consist of two polynomials each ($2n$ polynomials total). DRAMatic performs $4n$ sub-polynomial NTTs (because the polynomials are in RNS format), while $\text{MHY}^+$ instead performs $2n$ whole-polynomial NTTs.
When transforming just a single ciphertext, we see that the two implementations are similar in
pure computation time; $\text{MHY}^+$ takes 40 ms, while DRAMatic takes 42 ms.
However, DRAMatic's transfer and retrieval overhead is 36 times lower at 0.5 ms
compared to $\text{MHY}^+$'s 18 ms.
If we include this transfer and retrieval overhead, then
DRAMatic is 1.4 times faster than $\text{MHY}^+$, even for a single ciphertext.
The reason for this difference is that DRAMatic offloads the whole NTT operation at once,
while $\text{MHY}^+$ instead offloads individual NTT stages.

A bigger difference arises when transforming multiple ciphertexts at once.
$\text{MHY}^+$ uses all available DPUs even for a single ciphertext
and its computation time thus rises linearly with the number of ciphertexts
to be transformed.
DRAMatic instead performs the whole operation on individual DPUs and can thus
assign the additional ciphertexts to new DPUs that were previously unutilized.
As such, the computation time for DRAMatic stays low as the number of ciphertexts increases,
until the 128 DPUs are fully utilized.
As a result, DRAMatic gains more of an advantage
when more ciphertexts can be processed in parallel.
With four ciphertexts, DRAMatic is already 3.8 times faster in pure computation time.
And for 64 ciphertexts, DRAMatic is 60.6 times faster than $\text{MHY}^+$.
This scaling levels out at 512 ciphertexts when all 128 DPUs are fully utilized,
as shown in \autoref{fig:plot-vs-unine-multiple} (right).
For 512 ciphertexts and above, DRAMatic is
about 334 times faster in pure computation time.
The transfer and retrieve times also scale better for DRAMatic
and are about 630 times faster than $\text{MHY}^+$ for large numbers of ciphertexts.
In relative terms, compared to the pure computation time, $\text{MHY}^+$ has transfer and retrieve overheads of about 25\% and 20\% respectively, while DRAMatic reduces these overheads
to only 7\% and 16\% respectively.
If we include these transfer and retrieval overheads, then
DRAMatic is about 380 times faster than $\text{MHY}^+$ for 512 ciphertexts and above.

We can see that the power consumption of $\text{MHY}^+$ stays at a constant 278 watts during these tests, which is expected, since $\text{MHY}^+$ runs at fixed parallelism and only the amount of work (number of ciphertexts) is changed.
DRAMatic instead slightly increases its power consumption from 311 watts up to 323 watts as the number of ciphertexts (and thus DPU utilization) increases.
Interestingly, the power consumption is always slightly higher for DRAMatic than for $\text{MHY}^+$, even though DRAMatic scales its DPU usage while $\text{MHY}^+$ always uses all DPUs.
One reason for this could be that $\text{MHY}^+$'s synchronization overhead causes more idle time for DPUs or other system components, which lowers their power consumption.
Nevertheless, even though DRAMatic has slightly higher power consumption,
it is still more energy efficient than $\text{MHY}^+$, because its execution time is significantly shorter.
For 512 ciphertexts and above, DRAMatic uses
about 284 times less energy than $\text{MHY}^+$ for pure computation and about 330 times less energy when including data transfer and retrieval.

DRAMatic also scales better with available parallelism than $\text{MHY}^+$.
When using 256 DPUs instead of 128, the pure computation time for NTT with 4096 ciphertexts drops by 50\% for DRAMatic. If we account for the transfer and retrieval overhead, the total time still drops by 46\%.
In contrast, $\text{MHY}^+$ cannot effectively use the additional DPUs and even becomes 8\% slower due to additional transfer and retrieval overhead.
DRAMatic's performance roughly doubles again if we use all 509 DPUs available on our PIM test machine, while $\text{MHY}^+$ is limited to 256 DPUs, because it requires the number of DPUs to be a power of two.

\subsection{Comparing with Microsoft SEAL}
We now evaluate the performance of DRAMatic compared to Microsoft SEAL \cite{sealcrypto}, an optimized HE implementation for CPUs, regarding both runtime and energy efficiency. Our PIM test machine is the same as above.
Our SEAL test machine is equipped with an Intel\textsuperscript{\textregistered} Xeon\textsuperscript{\textregistered} Gold 5415+ CPU and 128 GB of DDR5 memory.
On this machine, we run SEAL version 4.1.2 compiled with clang-12 using the project's default CMake Release target.
We also measure the power consumption of the SEAL machine using a GUDE Expert Power Control 8035 metered PDU, which again measures full system power.


\subsection{Scaling with Polynomial Length}
\begin{figure}[t]
    \centering
    \includegraphics[width=0.94\linewidth]{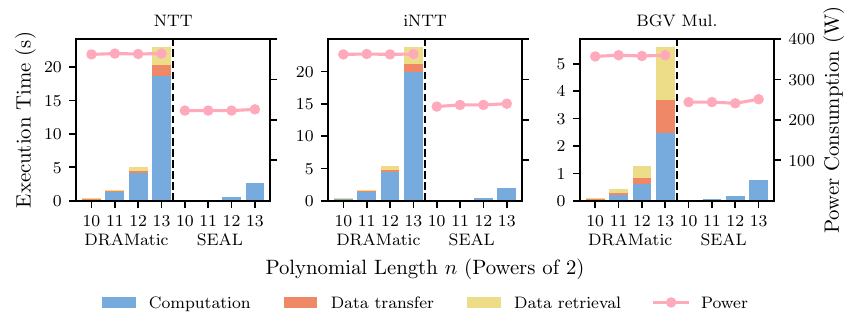}
    \caption{Performance of DRAMatic (DPU) compared to SEAL (CPU).}
    \label{fig:plot-op-multiple}
\end{figure}

\autoref{fig:plot-op-multiple} compares DRAMatic and SEAL for NTT, iNTT, and BGV multiplication across polynomial lengths from 1024 up to 8192.
The remaining parameters are as follows: 6 DPU ranks for DRAMatic (383 DPUs, parallel implementation) / 16 threads for SEAL, 32768 ciphertexts, and coefficient sizes per the Homomorphic Encryption Standard (27-, 54-, 109-, 218-bit) to achieve 128-bit security \cite{HomomorphicEncryptionSecurityStandard}.
Note that due to BGV's modulus switching, these coefficients can also be smaller for some operations.
For BGV multiplication, pairs of ciphertexts are multiplied to produce a single result ciphertext for each pair; inputs and outputs remain in NTT-form.
The raw results are listed in \autoref{sec:appendix-raw-results} (\autoref{tab:raw-results-8}).

On the left of \autoref{fig:plot-op-multiple},
we can see that for both implementations, the NTT execution time increases quadratically with longer polynomial lengths and the correspondingly larger coefficients.
However, SEAL performs better in general and is about 7 times
faster than DRAMatic in pure computation time.
Compared to $\text{MHY}^+$, this is a significant improvement.
$\text{MHY}^+$ is about 8600 times slower than SEAL in supported configurations (polynomial length of 2048), while DRAMatic closes the gap and is only about 7 times slower than SEAL in this test (using 383 DPUs).
We also measure the time required to transfer and retrieve the test data and results from the DPUs. In this test, the transfer and retrieval overhead of DRAMatic
is about 7\% and 16\% respectively.
Note however that in a real use-case, the data retrieval step is not necessary directly after an NTT, as the result would typically be the basis for further computations, and only the final results would be retrieved.
If the host actually requires the result of an NTT, it should compute the NTT itself, because, as we have seen, SEAL is still faster here.
A more realistic scenario with combined NTT, BGV multiplication, and iNTT is shown below.
We can also see that the power consumption of both systems stays almost constant during these tests, which is expected, since they both run at fixed parallelism and only the amount of work (length of polynomials and coefficients) is changed.
However, the PIM system uses about 62\% more power than the SEAL system in these tests.
If we account for the computation time, this shows that DRAMatic consumes about 10 times more energy than SEAL for the tested NTT operations, a significant decrease from $\text{MHY}^+$, which required about 10700 times more energy than SEAL in supported configurations.

The iNTT results (in the middle of \autoref{fig:plot-op-multiple}) are very similar to the NTT results.
However, the BGV multiplication tests (on the right of \autoref{fig:plot-op-multiple}) show some differences.
Notice that the data transfer and retrieval steps now take up a much larger share of the total execution time on DPUs. Their overhead compared to the pure computation is about 42\% and 73\% respectively.
This indicates that BGV multiplication has lower arithmetic intensity than NTT/iNTT.
DRAMatic also performs better in this test and is only about 3 times slower than SEAL in pure computation time.
As the power consumption is similar to before, DRAMatic's energy consumption in this test is thus also only 4.8 times higher than SEAL's.
Execution time still scales quadratically with polynomial length and the corresponding coefficient sizes.

\subsection{Scaling with the Number of Ciphertexts} \label{sec:scaling-with-nr-ciphertexts}
We now test a combined NTT, BGV multiplication, and iNTT pipeline for different numbers of ciphertexts.
The remaining parameters are as follows: 6 DPU ranks for DRAMatic (383 DPUs, parallel implementation) / 16 threads for SEAL, polynomial length 4096, and 109-bit coefficients.
The inputs start in non-NTT form, and the outputs are also transformed back into non-NTT form. For DRAMatic, transfers occur only before NTT and after iNTT.

The raw results are listed in \autoref{sec:appendix-raw-results} (\autoref{tab:raw-results-8.5}).
Both DRAMatic and SEAL show predictable linear scaling in execution time as the number of ciphertexts increases.
In this combined NTT, BGV multiplication, and iNTT test, we can now take a closer look at DRAMatic's data transfer and retrieval overhead; they are reduced to about 3\% and 6\% respectively.
In pure computation time, DRAMatic is about 6 times
slower than SEAL in these tests. Accounting for the data transfer and retrieval overhead, it is about 7 times slower.
Additionally, the SEAL system still uses about 38\% less power than the PIM system in these tests.

\subsection{Scaling with Available Parallelism} \label{sec:scaling-with-parallelism}
To determine how DRAMatic scales with available parallelism compared to SEAL, we limit the number of threads on the SEAL system and the number of DPUs on the PIM system in proportional steps.
We again test a combined NTT, BGV multiplication, and iNTT pipeline.
The remaining parameters are as follows: 32768 ciphertexts, polynomial length 4096, and 109-bit coefficients.
Again, inputs and outputs are in non-NTT form; and for DRAMatic, transfers occur only before NTT and after iNTT.
The raw results are listed in \autoref{sec:appendix-raw-results} (\autoref{tab:raw-results-9}).

\begin{figure}[t]
    \centering
    \includegraphics[width=0.94\linewidth]{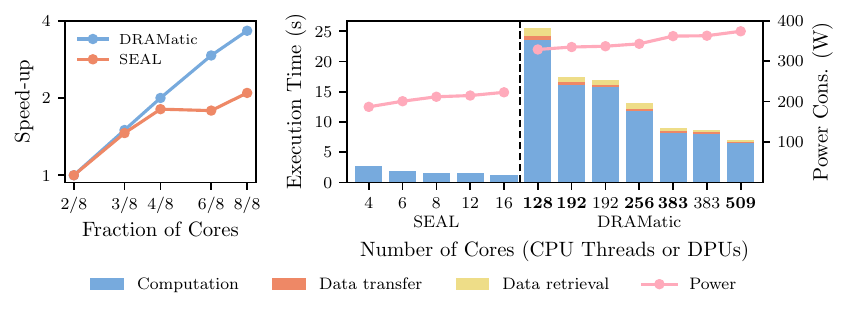}
    \caption{Combined NTT, BGV multiplication, and iNTT performance of DRAMatic (DPU) compared to SEAL (CPU) for different levels of parallelism. Bold data points (right) use DRAMatic's sequential implementation. The log-log plot (left) shows the speed-up of both implementations with increasing parallelism (computation time only).}
    \label{fig:plot-combined-multiple}
\end{figure}

The 109-bit coefficients in these tests require 3 RNS moduli.
Recall that DPU groups can only be allocated in whole ranks. Thus, DRAMatic's normal (parallel) implementation can only utilize either 3 or 6 of the 8 available DPU ranks.
To provide additional data points at 2, 4 and 8 ranks, we also employ the alternative, sequential implementation of DRAMatic (see \autoref{sec:splitting-over-DPUs}).
The sequential implementation is marked as bold in the results, which are shown in \autoref{fig:plot-combined-multiple} (right).
We can see that the sequential implementation of DRAMatic is slightly less efficient than our parallel implementation for numbers of DPUs which both can utilize (192 and 383).
However, this difference is quite small and the sequential implementation provides good approximations for the available performance with 128, 256, and 509 DPUs.
Note that only 509 of the 512 total DPUs are usable, as some DPU ranks of our test machine contain defective DPUs.

We can see that the execution times of both systems scale effectively with an increasing number of cores, while the power consumption moderately increases.
This is a significant advantage compared to $\text{MHY}^+$, which cannot effectively use more than 128 DPUs for NTT,
and also compared to concurrent work \cite{2026-NTT-UPMEM-PIM}, which is limited to 256 DPUs.
Note that for the SEAL system, the scaling effectivity becomes weaker once the number of threads exceeds the number of physical cores (8) and the system starts relying on simultaneous multi-threading (SMT).
In the log-log speed-up plot (on the left of \autoref{fig:plot-combined-multiple}, showing pure computation time, using DRAMatic's parallel implementation when available), we can see that DRAMatic scales more effectively with increasing parallelism, staying close to the theoretically optimal straight diagonal line. In contrast, SEAL's speed-up graph drops faster in these tests, especially once relying on SMT.

Additionally, the relative increase in power consumption is lower for the PIM system than for the SEAL system.
When going from 2 DPU ranks at 329 watts to 8 DPU ranks at 374 watts,
the PIM system becomes 3.7 times faster (in computation time), while only consuming 14\% more power.
The SEAL system, when going from 4 threads to 16 threads, becomes only about 2.1 times faster, but increases its power consumption by 19\%
from 187 watts to 223 watts.
This difference highlights the high baseline power consumption of the PIM system, which uses about 310 watts even when idle.
One reason for this could be the limiting of CPU C-states, which is a step that the UPMEM documentation lists as part of the server installation \cite{UPMEMdoc-Cstate}.
We suspect that this step might be required for stability reasons
and that the energy efficiency could potentially be improved in future UPMEM products.
The limiting of CPU C-states also suggests that UPMEM PIM could be more energy efficient for larger systems with more DPU ranks per CPU or if the CPU is simultaneously used for other tasks.

\section{Boosting Performance with Hardware Extensions}
We now discuss what hardware limitations are constraining DRAMatic's performance
and what hardware extensions could enable it to further catch up with optimized CPU implementations like SEAL.

\subsection{The Multiplication Bottleneck}
\begin{figure}[t]
    \centering
    \includegraphics[width=0.94\linewidth]{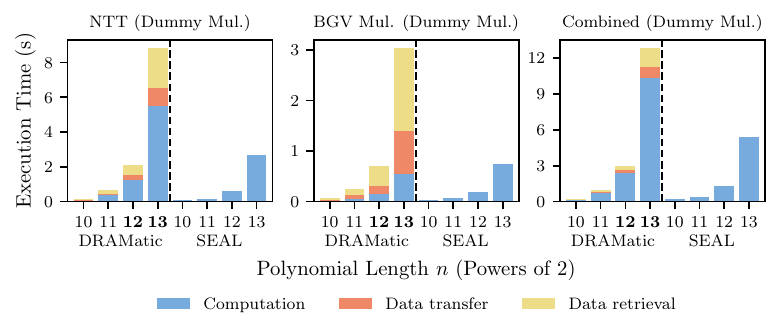}
    \caption{Simulated performance of DRAMatic using dummy multiplications (DPU) compared to SEAL (CPU). Bold data points use DRAMatic's sequential implementation.}
    \label{fig:plot-dummy-mul-multiple}
\end{figure}

A major bottleneck for DRAMatic is the weak multiplication performance of DPUs.
The DPU hardware only supports 8x8-bit multiplications and bigger multiplications need to be realized in software.
Although we leverage RNS representations to minimize the size of datatypes and implement optimized multiplication routines, the coefficient multiplications remain expensive on DPUs.
Even our most optimized 32-bit multiplication (with a 32-bit result) takes 21 cycles on a DPU. For comparison, modern CPUs can perform these multiplications in a single cycle, or even perform multiple such operations at the same time using instruction-level parallelism or SIMD instructions.

To gauge potential performance improvements from faster multiplications on DPUs, we run additional tests in which we replace DRAMatic's multiplication with fast dummy routines, which are about 9 to 10 times faster.
We test NTT; BGV multiplication; and a combined NTT, BGV multiplication, and iNTT pipeline across polynomial lengths from 1024 up to 8192.
The remaining parameters are as follows: 8 DPU ranks for DRAMatic (509 DPUs, parallel implementation when available) / 16 threads for SEAL, 32768 ciphertexts, and coefficient sizes per the Homomorphic Encryption Standard (27-, 54-, 109-, 218-bit) to achieve 128-bit security \cite{HomomorphicEncryptionSecurityStandard}.
The raw results are listed in \autoref{sec:appendix-raw-results} (\autoref{tab:raw-results-10}).

Our dummy 32-bit to 32-bit multiplication computes $(a + b) \oplus 12345678_{16}$ in 2 cycles, instead of the 21 cycles needed for proper multiplication.
And our dummy 32-bit to 64-bit multiplication computes $a \cdot 2^{20} + b$ in 4 cycles, instead of the normal 35 cycles.
For NTT, the dummy multiplication brings a 2.7 times speed-up. With this fast dummy multiplication, DRAMatic would thus only be around 2.2 times slower than SEAL in pure computation, as shown in \autoref{fig:plot-dummy-mul-multiple} (left).
The iNTT tests behave similarly, achieving a 2.8 times speed-up.
Of the tested operations, BGV multiplication benefits the most from these faster dummy multiplications. It achieves a 3.3 times speed-up and is thus even 1.3 times faster than SEAL in pure computation, as shown in \autoref{fig:plot-dummy-mul-multiple} (middle).
Note that the data transfer and retrieval overheads stay the same as in earlier tests and thus become more significant as the pure computation time decreases.
In the BGV multiplication test, the transfer and retrieval overheads dominate the total runtime at 150\% and 270\% respectively.
In the combined NTT, BGV multiplication, and iNTT tests, which depict these overheads in a more realistic scenario, the transfer and retrieval overhead are reduced to 10\% and 18\% respectively, as shown in \autoref{fig:plot-dummy-mul-multiple} (right).
The total speed-up due to the dummy multiplications is about 2.8 times in the combined test, as the NTT and iNTT steps outweigh the BGV multiplication step. DRAMatic would thus only be about 1.9 times slower than SEAL in the combined test.

We also test more optimistic dummy routines that simulate multiplications (32-bit to 32-bit and 32-bit to 64-bit) in only one clock cycle, closer to the speed of modern CPUs\footnote{Even if DPUs could multiply in one cycle, their clock frequency (400 MHz) is still lower than that of modern CPUs (multiple GHz).}.
In reality, these dummy routines compute $a + b$ and $(a+b) \cdot 2^{32} + a$ respectively and are inlined to prevent function call overhead.
When testing with these more optimistic dummy routines, the gap between DRAMatic and SEAL almost completely closes.
For NTT, the more optimistic dummy routines bring another 1.8 times speed-up, making DRAMatic only 1.2 times slower than SEAL in pure computation.
Similarly for iNTT, we would achieve an additional 1.7 times speed-up.
For BGV multiplication, the speed-up is 1.8 times, which would make DRAMatic 2.8 times faster than SEAL in pure computation.
In the combined NTT, BGV multiplication, and iNTT tests, the gap in computation time is almost closed, as DRAMatic remains only 7\% slower than SEAL.
However, as the data transfer and retrieval overheads stay the same, they become even more significant compared to the lower computation times.
For NTT, these overheads are 32\% and 65\% respectively. For iNTT, they are 30\% and 62\%.
For BGV multiplication, they are 285\% and 436\%. And for combined NTT, BGV multiplication, and iNTT, the overheads are 17\% and 30\%.

Note that these tests are only meant to demonstrate potential speed-ups due to faster multiplications on DPUs. They are based on dummy multiplication routines and cannot be used in a real scenario.
However, these tests show that the weak multiplication performance of DPUs remains a major bottleneck for DRAMatic
and that faster multiplication hardware on DPUs could bring great performance improvements and close the remaining gap between DRAMatic and SEAL.
These tests also show that HE on UPMEM PIM is compute-bound rather than memory-bound.

\subsection{Reducing Data Transfer Overhead}
Even though DRAMatic reduces the data transfer overhead about 630 times compared to previous work \cite{DBLP:Unine-UPMEM-HE},
it is still a significant cost, especially if the DPUs' multiplication performance were to improve as simulated above.
While DRAMatic's normal transfer and retrieval overhead is about 9\% total in a combined NTT, BGV multiplication, and iNTT test (\autoref{fig:plot-combined-multiple}),
it rises to 28\% if we simulate improved DPU multiplication performance (right of \autoref{fig:plot-dummy-mul-multiple}), or even 47\% with the more optimistic dummy routines.
With the simulated DPU multiplication performance, DRAMatic could be 1.3 times faster than SEAL for pure BGV multiplication, were it not for the significant transfer and retrieval overhead of about 420\% total, as shown in \autoref{fig:plot-dummy-mul-multiple} (middle).
This is especially relevant in this scenario, since
we have some operations (NTT/iNTT) which are faster for SEAL,
and some operations (BGV multiplication) which are faster for DRAMatic (excluding transfer overhead).
If we could overcome the transfer overhead in this scenario, it would enable a form of \textit{hybrid computing} in which the CPU and DPUs work together and each perform the sub-tasks for which they are most suitable, improving overall performance.

However, this is currently not possible with UPMEM PIM,
as the CPU cannot operate normally on PIM DIMM memory.
Instead, the CPU must use normal DRAM for its computations and copy the data to the PIM DIMMs when work should be offloaded to DPUs.
This is required, in part, because the memory chips on DIMMs are interleaved, \ie, logically consecutive bytes are stored in separate DRAM-chips.
The data must thus be transposed when making it available to DPUs, as they can only access one DRAM-chip each. The UPMEM runtime does this transposition automatically, but it requires a copy.

UPMEM PIM is most suited for operations with large amounts of data and low arithmetic intensity, as it removes the bottleneck of getting the data from memory to the processing units.
However, if the data is dynamic, then these operations also have the most substantial copying overhead, as the data must be copied to the DPUs every time.
In this case, it is faster to perform the computations directly on the CPU, since the data must be read anyway to copy it.
To improve the performance of UPMEM PIM in these use-cases with frequently changing data, like HE, we propose a hybrid computing system, which allows the CPU to use the DPUs memory directly when the DPUs are not currently active.
This hybrid system would allow DPUs to immediately continue a computation whose previous steps were performed by the CPU, as long as the CPU finishes its computation with the proper memory transposition. In many cases, this transposition could also be combined with the main computation, without requiring an additional pass over the data, or copying it.
This proposed change could greatly reduce the data transfer overhead and would make UPMEM PIM behave more like memory that can also compute, instead of a discrete accelerator that data must be transferred to.

\section{Suitability of other HE schemes}
DRAMatic is the first public implementation of BGV operations on UPMEM PIM, and BGV is the main focus of this work.
Yet many of our techniques also transfer to other schemes, like efficient NTT, or work distribution between DPUs.
This section discusses some of the potential challenges and advantages of other HE schemes on UPMEM PIM.

\subsection{BFV}
BFV \cite{C:Brakerski12,EPRINT:FanVer12} is similar to BGV, but its multiplication entails additional scaling and rounding operations. These operations are difficult to implement efficiently on UPMEM PIM,
since DPU division performance is even more limited than multiplication, and RNS division is also a lot more involved.
Additionally, floating-point optimizations \cite{RSA:HalPolSho19}, cannot be used on DPUs, as floating-point is not supported.

\subsection{CKKS}
CKKS \cite{AC:CKKS17} requires \textit{rescaling} operations, which consist of divisions and rounding on RNS integers.
Like with BFV, this is difficult to implement efficiently on UPMEM PIM.

\subsection{TFHE}
TFHE \cite{JC:CGGI20} could be promising for UPMEM PIM, as it uses smaller polynomials and coefficients than BGV, but more of them. For example, some ciphertext types in TFHE are multi-dimensional polynomial matrices, containing many comparatively small polynomials.
However, we have seen that DRAMatic and SEAL scale similarly with polynomial length, so smaller polynomials are not an inherent advantage for DRAMatic. Nevertheless, the larger number of sub-operations in TFHE may reduce batching requirements, which would benefit DRAMatic.

\section{Related Work}
We classify the related work into categories:
Works that use UPMEM PIM for accelerating HE or NTT,
works that use UPMEM PIM in other cryptographic contexts,
works that accelerate HE using other PIM systems,
and works that accelerate HE without using PIM.

\subsection{Using UPMEM PIM for Accelerating HE} \label{sec:related-work-unine}
Previous work \cite{DBLP:Unine-UPMEM-HE}, which we compared against in \autoref{sec:evaluation},
has also explored using UPMEM PIM for accelerating HE. They implement polynomial addition and multiplication on UPMEM PIM and integrate their implementation with the OpenFHE library \cite{ACM:OpenFHE} using its hardware abstraction layer (HAL).
However, this HAL-based integration offloads many small (sub)-operations to DPUs, instead of the larger overarching operation as a whole. Especially for NTT, this requires many copy operations between the DPUs and the host, as well as complex synchronization, as NTT stages are offloaded individually.
In this work, we instead implement NTT on the DPUs directly and offload the whole operation at once. This saves the expensive copying and synchronization operations and enables better parallelism and further DPU-centric optimizations.

Additionally, for some operations, $\text{MHY}^+$ only measure their performance on 16-bit coefficients, which are too small for practical HE use. In this work, we instead test with larger, more realistic coefficients sizes and also implement custom multiplication routines to speed-up operations on these larger coefficients.
The maximum coefficient size of $\text{MHY}^+$ is also limited to 64-bit,
which greatly constrains the operations that can be homomorphically evaluated,
as coefficient size is tied to the noise-budget and thus the maximum multiplicative depth of HE operations (or bootstrapping frequency for FHE).
According to the Homomorphic Encryption Standard \cite{HomomorphicEncryptionSecurityStandard}, this maximum coefficient size corresponds to a polynomial length of 2048 for 128-bit security.
However, most non-trivial homomorphic evaluations require larger parameters.
For example, in homomorphic matrix multiplications \cite{CCS:JKLS18, DBLP:MatrixMul-HE} the matrix sizes are limited by the number of \textit{slots} which is proportional to the polynomial length.
With n=2048, previous work \cite{DBLP:Unine-UPMEM-HE} is thus limited to square matrix dimensions of 32,
while DRAMatic can support larger matrices.
As another example, Homomorphic Evaluation of the AES Circuit \cite{C:GenHalSma12}
requires modulus sizes of 493 bits with bootstrapping or 886 bits for all ten rounds without bootstrapping.
Such a computation thus cannot be performed with MHY+, as it is limited to 64-bit coefficients.
In contrast, DRAMatic does support these parameters and can be used to securely perform these non-trivial homomorphic evaluations.

\subsection{Using UPMEM PIM for HE-suitable NTT} \label{sec:related-work-2026}
Concurrent work \cite{2026-NTT-UPMEM-PIM}, which we will refer to as $\textit{BMP}^\textit{+}$, has also investigated optimized NTT implementations on UPMEM PIM, specifically for large parameters as required by some FHE use-cases.
They compare different modular reductions (Barret and Montgomery), radices, and NTT algorithms, which they call 1D NTT, 2D NTT, and 4D NTT.
But unlike DRAMatic, they do not implement other polynomial operations or HE multiplication (\eg, BGV).
DRAMatic's NTT implementation is similar to their 1D NTT approach (using Barret reduction).

Unfortunately, the implementation of $\text{BMP}^+$ is currently not publicly available.
However, we attempt to recreate the test conditions from their work ($2^{16}$ polynomial length, 1241-bit coefficients, and 256 DPUs).
With DRAMatic's NTT implementation, this takes about 32.6 ms per polynomial (averaged over 4096 polynomials that are transformed in parallel, including data transfer and retrieval overhead).
For comparison, the 1D NTT\_v2 Barret variant of $\text{BMP}^+$ takes 43.2 ms per polynomial, so DRAMatic's 1D NTT implementation is about 33\% faster.
However, their alternative 2D NTT Barret variant only takes 25.3 ms per polynomial,
which would be 29\% faster than DRAMatic, as DRAMatic does not currently support 2D NTT.
Their 4D NTT takes between 28.1 ms and 20.5 ms per polynomial based on the radix used,
so it could be up to 59\% faster than DRAMatic's 1D NTT in this use case.
They also present NTT variants that use Montgomery reduction and achieve further improvements,
but these are not directly comparable with DRAMatic, as they require pre-transforming the values into Montgomery space.

In addition to these differences, $\text{BMP}^+$ also requires specific DPU counts\footnote{In their journal pre-proof, they state that they selected 256 DPUs, as otherwise not all of their kernels can be implemented.},
while DRAMatic is more flexible and can utilize both larger and smaller DPU installations.
For example, using all 509 DPUs of our PIM test machine, DRAMatic's NTT time drops to 23 ms per polynomial, comparable to $\text{BMP}^+$'s 4D NTT that is limited to 256 DPUs.
Lastly, due to DRAMatic's coarse-grained multi-threading for NTT, it requires larger workloads (multiple ciphertexts) to reach its full performance (see for example \autoref{fig:plot-vs-unine-multiple}), while $\text{BMP}^+$ can also efficiently operate on single ciphertexts/polynomials.
In the future, it could be interesting to investigate whether DRAMatic's optimizations can be combined with the 2D or 4D NTT approach of $\text{BMP}^+$.

\subsection{Using UPMEM PIM for other Cryptography}
A different approach to secure computing on UPMEM PIM was investigated by \cite{ghinani2025}.
Instead of using HE like in this work, they leverage multi-party computation (MPC) techniques to integrate PIM modules within a secure computing system by protecting the data as it moves off-chip to the PIM modules.

\subsection{Accelerating HE using other PIM Systems}
In \cite{DBLP:Gupta2021-HEcustomPIMchips}, the authors design custom PIM hardware for accelerating HE, and evaluate its performance using a simulator, which shows a large theoretical speed-up compared to CPU implementations.
This is fundamentally different from our work, since they evaluate custom hardware designs and simulate them, while we instead implement and evaluate HE operations on real general-purpose PIM hardware.

\subsection{Accelerating HE without PIM}
Lastly, there has also been extensive research into accelerating HE on other computing architectures, separate from PIM. There are popular CPU implementations like OpenFHE \cite{ACM:OpenFHE} and Microsoft SEAL \cite{sealcrypto}, which we also compared against,
as well as optimized implementations for GPUs \cite{TCHES:JKACL21} and FPGAs \cite{CHES:PNPM15}.

\section{Conclusion}
We presented DRAMatic, which implements efficient polynomial operations and NTT
directly on UPMEM PIM DPUs, as the basis for HE operations.
DRAMatic supports large parameter sizes, as required for practical, secure homomorphic evaluations.
We addressed the unique challenges of the UPMEM PIM platform by implementing DPU-centric optimizations,
like custom multiplication routines, and by considering efficient data layouts.
Our results show promising scaling with increasing parallelism
and a 334 times speed-up compared to previous work on accelerating HE with UPMEM PIM \cite{DBLP:Unine-UPMEM-HE}, significantly closing the gap to optimized CPU implementations.
Our experiments revealed that the weak multiplication performance of DPUs
remains a bottleneck, and that on UPMEM PIM, HE is compute-bound rather than memory-bound.
We thus proposed hardware extensions to UPMEM PIM, which could eliminate these bottlenecks and further improve DRAMatic's performance.
Compared to concurrent work on accelerating NTT execution on UPMEM PIM \cite{2026-NTT-UPMEM-PIM}, we showed that our NTT implementation scales better with more DPUs and is 33\% faster than the directly comparable 1D NTT.
However, DRAMatic does not currently support 2D NTT or 4D NTT, which show even better performance.
Lastly, we measured and evaluated the power consumption of our test systems
and showed that the UPMEM PIM system is less energy efficient than the CPU system.
However, the scaling of power consumption with increasing parallelism looks
promising for UPMEM PIM and could benefit larger PIM installations.

\section{Future Work}
Future research could focus on integrating our proposed hardware extensions and hybrid computing architecture into UPMEM PIM or similar processing-in-memory platforms. A key area of interest is the potential synergy between DRAMatic's optimizations and recent 2D/4D NTT approaches \cite{2026-NTT-UPMEM-PIM} to further enhance throughput.
Achieving efficient bootstrapping remains a critical milestone to enable full FHE support on UPMEM PIM. Finally, investigating the portability of these optimizations to other programmable PIM architectures, such as AxDIMM \cite{DBLP:AxDIMM}, will provide insights into the broader applicability of our acceleration techniques.

\section*{Acknowledgments}
Generative AI was utilized to generate plots,
for editing, and for grammar enhancement of this work. This work has been supported by the BMFTR through the project AnoMed.

\bibliography{abbrev3,crypto,others}

@string{ieee =                  {IEEE}}

@string{acm =                   "{ACM}"}

@article{CooleyTukey1965-FFT,
  title={An algorithm for the machine calculation of complex {Fourier} series},
  author={Cooley, James W. and Tukey, John W.},
  journal={Mathematics of computation},
  volume={19},
  number={90},
  pages={297--301},
  year={1965},
  publisher={JSTOR},
  doi={10.2307/2003354}
}

@inproceedings{GentlemanSande1966-FFT,
  title={Fast {Fourier} transforms: for fun and profit},
  author={Gentleman, W. Morven and Sande, Gordon},
  booktitle={Proceedings of the November 7-10, 1966, fall joint computer conference},
  pages={563--578},
  year={1966},
  doi={10.1145/1464291.1464352}
}

@article{DBLP:GomezMutlu2021-BenchmarkingPIM,
  author       = {Juan G{\'{o}}mez{-}Luna and
                  Izzat El Hajj and
                  Ivan Fernandez and
                  Christina Giannoula and
                  Geraldo F. Oliveira and
                  Onur Mutlu},
  title        = {Benchmarking a New Paradigm: An Experimental Analysis of a Real Processing-in-Memory Architecture},
  journal      = {CoRR},
  volume       = {abs/2105.03814},
  year         = {2021},
  url          = {https://arxiv.org/abs/2105.03814},
  eprinttype   = {arXiv},
  eprint       = {2105.03814},
  bibsource    = {dblp computer science bibliography, https://dblp.org}
}

@book{Shoup2005-NumberTheoryAlgebra,
  place={Cambridge},
  title={A Computational Introduction to Number Theory and Algebra},
  publisher={Cambridge University Press},
  author={Shoup, Victor},
  year={2005}
}

@misc{sealcrypto,
  title = {{M}icrosoft {SEAL} (release 4.1)},
  howpublished = {\url{https://github.com/Microsoft/SEAL}},
  month = jan,
  year = 2023,
  note = {Microsoft Research, Redmond, WA.},
  key = {SEAL}
}

@techreport{HomomorphicEncryptionSecurityStandard,
  author = {Martin Albrecht and Melissa Chase and Hao Chen and Jintai Ding and Shafi Goldwasser and Sergey Gorbunov and Shai Halevi and Jeffrey Hoffstein and Kim Laine and Kristin Lauter and Satya Lokam and Daniele Micciancio and Dustin Moody and Travis Morrison and Amit Sahai and Vinod Vaikuntanathan},
  title = {Homomorphic Encryption Security Standard},
  institution= {HomomorphicEncryption.org},
  publisher = {HomomorphicEncryption.org},
  address = {Toronto, Canada},
  year = {2018},
  month = {November}
}

@misc{UPMEMdoc-Cstate,
  title = {{UPMEM} {DPU} {SDK} 2025.1.0 Documentation --- Server installation},
  howpublished = {\url{https://sdk.upmem.com/2025.1.0/280_Server_configuration.html}},
  key = {UPMEM},
  year = 2025
}

@misc{UPMEM2024-Keynote,
  author       = {UPMEM},
  title        = {Keynote: {UPMEM} {PIM} platform for Data-Intensive Applications},
  howpublished = {Presentation slides, ABUMPIMP 2024 at Euro-Par 2024},
  year         = {2024}
}

@inproceedings{DBLP:Gupta2021-HEcustomPIMchips,
  author       = {Saransh Gupta and
                  Tajana Simunic Rosing},
  title        = {Invited: Accelerating Fully Homomorphic Encryption with Processing
                  in Memory},
  booktitle    = {58th {ACM/IEEE} Design Automation Conference, {DAC} 2021, San Francisco,
                  CA, USA, December 5-9, 2021},
  pages        = {1335--1338},
  publisher    = {{IEEE}},
  year         = {2021},
  url          = {https://doi.org/10.1109/DAC18074.2021.9586285},
  doi          = {10.1109/DAC18074.2021.9586285},
  bibsource    = {dblp computer science bibliography, https://dblp.org}
}

@inproceedings{DBLP:Unine-UPMEM-HE,
  author       = {Mpoki Mwaisela and
                  Joel Hari and
                  Peterson Yuhala and
                  J{\"{a}}mes M{\'{e}}n{\'{e}}trey and
                  Pascal Felber and
                  Valerio Schiavoni},
  title        = {Evaluating the Potential of In-Memory Processing to Accelerate Homomorphic
                  Encryption: Practical Experience Report},
  booktitle    = {43rd International Symposium on Reliable Distributed Systems, {SRDS}
                  2024, Charlotte, NC, USA, September 30 - Oct. 3, 2024},
  pages        = {92--103},
  publisher    = {{IEEE}},
  year         = {2024},
  url          = {https://doi.org/10.1109/SRDS64841.2024.00019},
  doi          = {10.1109/SRDS64841.2024.00019},
  bibsource    = {dblp computer science bibliography, https://dblp.org}
}

@article{Ghose2019-PIM-Perspective,
  author       = {Saugata Ghose and
                  Amirali Boroumand and
                  Jeremie S. Kim and
                  Juan G{\'{o}}mez{-}Luna and
                  Onur Mutlu},
  title        = {Processing-in-memory: {A} workload-driven perspective},
  journal      = {{IBM} J. Res. Dev.},
  volume       = {63},
  number       = {6},
  pages        = {3:1--3:19},
  year         = {2019},
  url          = {https://doi.org/10.1147/JRD.2019.2934048},
  doi          = {10.1147/JRD.2019.2934048},
  bibsource    = {dblp computer science bibliography, https://dblp.org}
}

@article{Zhou2025-FHEmem-Accelerator,
  author={Zhou, Minxuan and Nam, Yujin and Gangwar, Pranav and Xu, Weihong and Dutta, Arpan and Wilkerson, Chris and Cammarota, Rosario and Gupta, Saransh and Rosing, Tajana},
  journal={IEEE Transactions on Emerging Topics in Computing}, 
  title={{FHEmem}: A Processing In-Memory Accelerator for Fully Homomorphic Encryption}, 
  year={2025},
  volume={},
  number={},
  pages={1-16},
  doi={10.1109/TETC.2025.3528862}
}

@article{Satriawan2023-Review-on-NTT,
  author={Satriawan, Ardianto and Syafalni, Infall and Mareta, Rella and Anshori, Isa and Shalannanda, Wervyan and Barra, Aleams},
  journal={IEEE Access}, 
  title={Conceptual Review on Number Theoretic Transform and Comprehensive Review on Its Implementations}, 
  year={2023},
  volume={11},
  number={},
  pages={70288-70316},
  doi={10.1109/ACCESS.2023.3294446}
}

@article{DBLP:AxDIMM,
  author       = {Liu Ke and
                  Xuan Zhang and
                  Jinin So and
                  Jong{-}Geon Lee and
                  Shinhaeng Kang and
                  Sukhan Lee and
                  Songyi Han and
                  YeonGon Cho and
                  Jin Hyun Kim and
                  Yongsuk Kwon and
                  KyungSoo Kim and
                  Jin Jung and
                  IlKwon Yun and
                  Sung Joo Park and
                  Hyunsun Park and
                  Joon{-}Ho Song and
                  Jeonghyeon Cho and
                  Kyomin Sohn and
                  Nam Sung Kim and
                  Hsien{-}Hsin S. Lee},
  title        = {Near-Memory Processing in Action: Accelerating Personalized Recommendation
                  With {AxDIMM}},
  journal      = {{IEEE} Micro},
  volume       = {42},
  number       = {1},
  pages        = {116--127},
  year         = {2022},
  url          = {https://doi.org/10.1109/MM.2021.3097700},
  doi          = {10.1109/MM.2021.3097700},
  bibsource    = {dblp computer science bibliography, https://dblp.org}
}

@inproceedings{tee-fail,
    title     = {{TEE}.fail: Breaking Trusted Execution Environments via {DDR5} Memory Bus Interposition},
    booktitle = {47th IEEE Symposium on Security and Privacy},
    author    = {Jalen Chuang and Alex Seto and Nicolas Berrios and Stephan van Schaik and Christina Garman and Daniel Genkin},
    url       = {https://tee.fail},
    publisher = {IEEE Computer Society},
    year      = {2026},
}

@inproceedings{DBLP:MatrixMul-HE,
  author       = {Panagiotis Rizomiliotis and
                  Aikaterini Triakosia},
  editor       = {Francesco Regazzoni and
                  Marten van Dijk},
  title        = {On Matrix Multiplication with Homomorphic Encryption},
  booktitle    = {Proceedings of the 2022 on Cloud Computing Security Workshop, {CCSW}
                  2022, Los Angeles, CA, USA, 7 November 2022},
  pages        = {53--61},
  publisher    = {{ACM}},
  year         = {2022},
  url          = {https://doi.org/10.1145/3560810.3564267},
  doi          = {10.1145/3560810.3564267},
  bibsource    = {dblp computer science bibliography, https://dblp.org}
}

@article{2026-NTT-UPMEM-PIM,
title = {Long Integer {NTT} Execution on {UPMEM-PIM} for 128-bit Secure Fully Homomorphic Encryption},
journal = {Future Generation Computer Systems},
pages = {108386},
year = {2026},
issn = {0167-739X},
doi = {https://doi.org/10.1016/j.future.2026.108386},
url = {https://www.sciencedirect.com/science/article/pii/S0167739X26000208},
author = {Tathagata Barik and Priyam Mehta and Zaira Pindado and Harshita Gupta and Mayank Kabra and Mohammad Sadrosadati and Onur Mutlu and Antonio J. Peña}
}

@inproceedings{ghinani2025,
  author       = {Sahar Ghoflsaz Ghinani and
                  Jingyao Zhang and
                  Elaheh Sadredini},
  editor       = {Lujo Bauer and
                  Giancarlo Pellegrino},
  title        = {Enabling Low-Cost Secure Computing on Untrusted In-Memory Architectures},
  booktitle    = {34th {USENIX} Security Symposium, {USENIX} Security 2025, Seattle,
                  WA, USA, August 13-15, 2025},
  pages        = {1749--1767},
  publisher    = {{USENIX} Association},
  year         = {2025},
  url          = {https://www.usenix.org/conference/usenixsecurity25/presentation/ghinani},
  bibsource    = {dblp computer science bibliography, https://dblp.org}
}

@inproceedings{ACM:OpenFHE,
author = {Al Badawi, Ahmad and Bates, Jack and Bergamaschi, Flavio and Cousins, David Bruce and Erabelli, Saroja and Genise, Nicholas and Halevi, Shai and Hunt, Hamish and Kim, Andrey and Lee, Yongwoo and Liu, Zeyu and Micciancio, Daniele and Quah, Ian and Polyakov, Yuriy and R.V., Saraswathy and Rohloff, Kurt and Saylor, Jonathan and Suponitsky, Dmitriy and Triplett, Matthew and Vaikuntanathan, Vinod and Zucca, Vincent},
title = {{OpenFHE}: Open-Source Fully Homomorphic Encryption Library},
year = {2022},
isbn = {9781450398770},
publisher = {Association for Computing Machinery},
address = {New York, NY, USA},
url = {https://doi.org/10.1145/3560827.3563379},
doi = {10.1145/3560827.3563379},
booktitle = {Proceedings of the 10th Workshop on Encrypted Computing \& Applied Homomorphic Cryptography},
pages = {53–63},
numpages = {11},
location = {Los Angeles, CA, USA},
series = {WAHC'22}
}


\appendix
\section{Raw Results} \label{sec:appendix-raw-results}
This section contains the raw result tables.
Compute, Transfer, and Retrieve times are in seconds. Modulus length (Mod.) of the evaluation context is given in bits; it is often lower than the nominal (key context) length due to BGV's modulus switching. Power (Pwr.) is in Watts.
Bold data points (Cores) use DRAMatic's sequential implementation.

\begin{table}[ht]
\centering
{\fontsize{7pt}{8.4pt}\selectfont
\addtolength{\tabcolsep}{-1.6pt}
\begin{tabular}{llrrrrrrrr}
\hline
\textbf{Impl.} & \textbf{Operation} & \textbf{Cores} & \textbf{Degree} & \textbf{Mod.} & \textbf{C.texts} & \textbf{Compute} & \textbf{Transfer} & \textbf{Retrieve} & \textbf{Pwr.} \\
\hline
DRAMatic & NTT & 2 & 2048 & 54 & 1 & 0.042 & 0.0 & 0.0 & 311 \\
DRAMatic & NTT & 8 & 2048 & 54 & 4 & 0.042 & 0.0 & 0.0 & 311 \\
DRAMatic & NTT & 32 & 2048 & 54 & 16 & 0.042 & 0.0 & 0.0 & 311 \\
DRAMatic & NTT & 128 & 2048 & 54 & 64 & 0.042 & 0.001 & 0.001 & 313 \\
DRAMatic & NTT & 128 & 2048 & 54 & 256 & 0.042 & 0.003 & 0.005 & 320 \\
DRAMatic & NTT & 128 & 2048 & 54 & 512 & 0.061 & 0.005 & 0.009 & 323 \\
DRAMatic & NTT & 128 & 2048 & 54 & 1024 & 0.121 & 0.009 & 0.018 & 323 \\
DRAMatic & NTT & 128 & 2048 & 54 & 4096 & 0.485 & 0.035 & 0.068 & 323 \\
$\text{MHY}^+$ & NTT & 128 & 2048 & 54 & 1 & 0.040 & 0.010 & 0.008 & 278 \\
$\text{MHY}^+$ & NTT & 128 & 2048 & 54 & 4 & 0.159 & 0.040 & 0.032 & 278 \\
$\text{MHY}^+$ & NTT & 128 & 2048 & 54 & 16 & 0.636 & 0.160 & 0.129 & 278 \\
$\text{MHY}^+$ & NTT & 128 & 2048 & 54 & 64 & 2.542 & 0.638 & 0.512 & 278 \\
$\text{MHY}^+$ & NTT & 128 & 2048 & 54 & 256 & 10.024 & 2.241 & 1.823 & 278 \\
$\text{MHY}^+$ & NTT & 128 & 2048 & 54 & 512 & 20.048 & 4.481 & 3.648 & 278 \\
$\text{MHY}^+$ & NTT & 128 & 2048 & 54 & 1024 & 40.462 & 9.785 & 7.735 & 278 \\
$\text{MHY}^+$ & NTT & 128 & 2048 & 54 & 4096 & 160.542 & 36.138 & 29.131 & 278 \\
\hline
DRAMatic & iNTT & 2 & 2048 & 54 & 1 & 0.045 & 0.0 & 0.0 & 312 \\
DRAMatic & iNTT & 8 & 2048 & 54 & 4 & 0.045 & 0.0 & 0.0 & 312 \\
DRAMatic & iNTT & 32 & 2048 & 54 & 16 & 0.045 & 0.0 & 0.0 & 312 \\
DRAMatic & iNTT & 128 & 2048 & 54 & 64 & 0.045 & 0.001 & 0.001 & 314 \\
DRAMatic & iNTT & 128 & 2048 & 54 & 256 & 0.045 & 0.003 & 0.005 & 321 \\
DRAMatic & iNTT & 128 & 2048 & 54 & 512 & 0.065 & 0.005 & 0.009 & 322 \\
DRAMatic & iNTT & 128 & 2048 & 54 & 1024 & 0.130 & 0.009 & 0.018 & 322 \\
DRAMatic & iNTT & 128 & 2048 & 54 & 4096 & 0.519 & 0.023 & 0.054 & 325 \\
$\text{MHY}^+$ & iNTT & 128 & 2048 & 54 & 1 & 0.040 & 0.010 & 0.008 & 287 \\
$\text{MHY}^+$ & iNTT & 128 & 2048 & 54 & 4 & 0.159 & 0.039 & 0.033 & 287 \\
$\text{MHY}^+$ & iNTT & 128 & 2048 & 54 & 16 & 0.635 & 0.158 & 0.130 & 287 \\
$\text{MHY}^+$ & iNTT & 128 & 2048 & 54 & 64 & 2.532 & 0.630 & 0.523 & 287 \\
$\text{MHY}^+$ & iNTT & 128 & 2048 & 54 & 256 & 10.113 & 2.495 & 2.033 & 287 \\
$\text{MHY}^+$ & iNTT & 128 & 2048 & 54 & 512 & 20.236 & 5.011 & 4.062 & 287 \\
$\text{MHY}^+$ & iNTT & 128 & 2048 & 54 & 1024 & 40.427 & 9.982 & 8.032 & 287 \\
$\text{MHY}^+$ & iNTT & 128 & 2048 & 54 & 4096 & 161.704 & 39.972 & 32.239 & 287 \\
\hline
\end{tabular}
}
\caption{Raw performance results of \autoref{fig:plot-vs-unine-multiple}.}
\label{tab:raw-results-7}
\end{table}

\begin{table}
\centering
{\fontsize{7pt}{8.4pt}\selectfont
\addtolength{\tabcolsep}{-1.6pt}
\begin{tabular}{llrrrrrrrr}
\hline
\textbf{Impl.} & \textbf{Operation} & \textbf{Cores} & \textbf{Degree} & \textbf{Mod.} & \textbf{C.texts} & \textbf{Compute} & \textbf{Transfer} & \textbf{Retrieve} & \textbf{Pwr.} \\
\hline
DRAMatic & NTT & 383 & 1024 & 27 & 32768 & 0.296 & 0.028 & 0.065 & 362 \\
DRAMatic & NTT & 383 & 2048 & 54 & 32768 & 1.314 & 0.102 & 0.213 & 364 \\
DRAMatic & NTT & 383 & 4096 & 72 & 32768 & 4.130 & 0.258 & 0.705 & 363 \\
DRAMatic & NTT & 383 & 8192 & 174 & 32768 & 18.713 & 1.550 & 2.787 & 364 \\
SEAL & NTT & 16 & 1024 & 27 & 32768 & 0.092 & 0.0 & 0.0 & 223 \\
SEAL & NTT & 16 & 2048 & 54 & 32768 & 0.149 & 0.0 & 0.0 & 223 \\
SEAL & NTT & 16 & 4096 & 72 & 32768 & 0.631 & 0.0 & 0.0 & 223 \\
SEAL & NTT & 16 & 8192 & 174 & 32768 & 2.654 & 0.0 & 0.0 & 226 \\
\hline
DRAMatic & iNTT & 383 & 1024 & 27 & 32768 & 0.320 & 0.027 & 0.065 & 362 \\
DRAMatic & iNTT & 383 & 2048 & 54 & 32768 & 1.408 & 0.106 & 0.185 & 363 \\
DRAMatic & iNTT & 383 & 4096 & 72 & 32768 & 4.392 & 0.323 & 0.582 & 362 \\
DRAMatic & iNTT & 383 & 8192 & 174 & 32768 & 19.869 & 1.297 & 2.637 & 363 \\
SEAL & iNTT & 16 & 1024 & 27 & 32768 & 0.064 & 0.0 & 0.0 & 233 \\
SEAL & iNTT & 16 & 2048 & 54 & 32768 & 0.112 & 0.0 & 0.0 & 237 \\
SEAL & iNTT & 16 & 4096 & 72 & 32768 & 0.474 & 0.0 & 0.0 & 237 \\
SEAL & iNTT & 16 & 8192 & 174 & 32768 & 2.016 & 0.0 & 0.0 & 240 \\
\hline
DRAMatic & BGV mul. & 383 & 1024 & 27 & 32768 & 0.053 & 0.025 & 0.037 & 357 \\
DRAMatic & BGV mul. & 383 & 2048 & 54 & 32768 & 0.210 & 0.082 & 0.139 & 360 \\
DRAMatic & BGV mul. & 383 & 4096 & 72 & 32768 & 0.595 & 0.231 & 0.454 & 358 \\
DRAMatic & BGV mul. & 383 & 8192 & 174 & 32768 & 2.464 & 1.218 & 1.944 & 360 \\
SEAL & BGV mul. & 16 & 1024 & 27 & 32768 & 0.038 & 0.0 & 0.0 & 244 \\
SEAL & BGV mul. & 16 & 2048 & 54 & 32768 & 0.063 & 0.0 & 0.0 & 244 \\
SEAL & BGV mul. & 16 & 4096 & 72 & 32768 & 0.186 & 0.0 & 0.0 & 241 \\
SEAL & BGV mul. & 16 & 8192 & 174 & 32768 & 0.747 & 0.0 & 0.0 & 251 \\
\hline
DRAMatic & NTT & 383 & 16384 & 377 & 4096 & 13.547 & 0.703 & 1.101 \\
DRAMatic & NTT & 383 & 32768 & 812 & 4096 & 62.335 & 2.985 & 5.089 \\
SEAL & NTT & 16 & 16384 & 389 & 4096 & 1.397 & 0.0 & 0.0 \\
SEAL & NTT & 16 & 32768 & 825 & 4096 & 5.488 & 0.0 & 0.0 \\
\hline
DRAMatic & iNTT & 383 & 16384 & 377 & 4096 & 14.274 & 0.776 & 1.118 \\
DRAMatic & iNTT & 383 & 32768 & 812 & 4096 & 65.486 & 3.603 & 5.049 \\
SEAL & iNTT & 16 & 16384 & 389 & 4096 & 1.072 & 0.0 & 0.0 \\
SEAL & iNTT & 16 & 32768 & 825 & 4096 & 4.266 & 0.0 & 0.0 \\
\hline
DRAMatic & BGV mul. & 383 & 16384 & 377 & 4096 & 1.464 & 0.620 & 1.109 \\
DRAMatic & BGV mul. & 383 & 32768 & 812 & 4096 & 6.252 & 2.669 & 4.952 \\
SEAL & BGV mul. & 16 & 16384 & 389 & 4096 & 0.362 & 0.0 & 0.0 \\
SEAL & BGV mul. & 16 & 32768 & 825 & 4096 & 1.319 & 0.0 & 0.0 \\
\hline
\end{tabular}
}
\caption{Raw performance results of \autoref{fig:plot-op-multiple} and additional results for larger polynomial degrees.}
\label{tab:raw-results-8}
\end{table}

\begin{table}
\centering
{\fontsize{7pt}{8.4pt}\selectfont
\addtolength{\tabcolsep}{-1.6pt}
\begin{tabular}{llrrrrrrrr}
\hline
\textbf{Impl.} & \textbf{Operation} & \textbf{Cores} & \textbf{Degree} & \textbf{Mod.} & \textbf{C.texts} & \textbf{Compute} & \textbf{Transfer} & \textbf{Retrieve} & \textbf{Pwr.} \\
\hline
DRAMatic & Combined & 383 & 4096 & 72 & 8192 & 2.148 & 0.072 & 0.120 & 361 \\
DRAMatic & Combined & 383 & 4096 & 72 & 16384 & 4.111 & 0.120 & 0.267 & 362 \\
DRAMatic & Combined & 383 & 4096 & 72 & 32768 & 8.042 & 0.256 & 0.444 & 363 \\
SEAL & Combined & 16 & 4096 & 72 & 8192 & 0.326 & 0.0 & 0.0 & 223 \\
SEAL & Combined & 16 & 4096 & 72 & 16384 & 0.652 & 0.0 & 0.0 & 223 \\
SEAL & Combined & 16 & 4096 & 72 & 32768 & 1.291 & 0.0 & 0.0 & 223 \\
\hline
\end{tabular}
}
\caption{Raw performance results of \autoref{sec:scaling-with-nr-ciphertexts}.}
\label{tab:raw-results-8.5}
\end{table}

\begin{table}
\centering
{\fontsize{7pt}{8.4pt}\selectfont
\addtolength{\tabcolsep}{-1.6pt}
\begin{tabular}{llrrrrrrrr}
\hline
\textbf{Impl.} & \textbf{Operation} & \textbf{Cores} & \textbf{Degree} & \textbf{Mod.} & \textbf{C.texts} & \textbf{Compute} & \textbf{Transfer} & \textbf{Retrieve} & \textbf{Pwr.} \\
\hline
DRAMatic & Combined & \textbf{128} & 4096 & 72 & 32768 & 23.559 & 0.731 & 1.170 & 329 \\
DRAMatic & Combined & \textbf{192} & 4096 & 72 & 32768 & 16.126 & 0.494 & 0.829 & 335 \\
DRAMatic & Combined & 192 & 4096 & 72 & 32768 & 15.702 & 0.480 & 0.832 & 337 \\
DRAMatic & Combined & \textbf{256} & 4096 & 72 & 32768 & 11.775 & 0.390 & 0.895 & 343 \\
DRAMatic & Combined & \textbf{383} & 4096 & 72 & 32768 & 8.166 & 0.264 & 0.505 & 362 \\
DRAMatic & Combined & 383 & 4096 & 72 & 32768 & 8.042 & 0.256 & 0.444 & 363 \\
DRAMatic & Combined & \textbf{509} & 4096 & 72 & 32768 & 6.443 & 0.186 & 0.380 & 374 \\
SEAL & Combined & 4 & 4096 & 72 & 32768 & 2.703 & 0.0 & 0.0 & 187 \\
SEAL & Combined & 6 & 4096 & 72 & 32768 & 1.851 & 0.0 & 0.0 & 201 \\
SEAL & Combined & 8 & 4096 & 72 & 32768 & 1.494 & 0.0 & 0.0 & 212 \\
SEAL & Combined & 12 & 4096 & 72 & 32768 & 1.513 & 0.0 & 0.0 & 215 \\
SEAL & Combined & 16 & 4096 & 72 & 32768 & 1.291 & 0.0 & 0.0 & 223 \\
\hline
\end{tabular}
}
\caption{Raw performance results of \autoref{fig:plot-combined-multiple}.}
\label{tab:raw-results-9}
\end{table}

\begin{table}
\centering
{\fontsize{7pt}{8.4pt}\selectfont
\addtolength{\tabcolsep}{-1.6pt}
\begin{tabular}{llrrrrrrr}
\hline
\textbf{Impl.} & \textbf{Operation} & \textbf{Cores} & \textbf{Degree} & \textbf{Mod.} & \textbf{C.texts} & \textbf{Compute} & \textbf{Transfer} & \textbf{Retrieve} \\
\hline
DRAMatic & NTT & 509 & 1024 & 27 & 32768 & 0.088 & 0.021 & 0.053 \\
DRAMatic & NTT & 509 & 2048 & 54 & 32768 & 0.371 & 0.085 & 0.192 \\
DRAMatic & NTT & \textbf{509} & 4096 & 72 & 32768 & 1.266 & 0.260 & 0.576 \\
DRAMatic & NTT & \textbf{509} & 8192 & 174 & 32768 & 5.509 & 1.002 & 2.339 \\
SEAL & NTT & 16 & 1024 & 27 & 32768 & 0.092 & 0.0 & 0.0 \\
SEAL & NTT & 16 & 2048 & 54 & 32768 & 0.149 & 0.0 & 0.0 \\
SEAL & NTT & 16 & 4096 & 72 & 32768 & 0.631 & 0.0 & 0.0 \\
SEAL & NTT & 16 & 8192 & 174 & 32768 & 2.654 & 0.0 & 0.0 \\
\hline
DRAMatic & BGV mul. & 509 & 1024 & 27 & 32768 & 0.013 & 0.022 & 0.032 \\
DRAMatic & BGV mul. & 509 & 2048 & 54 & 32768 & 0.050 & 0.087 & 0.117 \\
DRAMatic & BGV mul. & \textbf{509} & 4096 & 72 & 32768 & 0.143 & 0.174 & 0.394 \\
DRAMatic & BGV mul. & \textbf{509} & 8192 & 174 & 32768 & 0.551 & 0.849 & 1.646 \\
SEAL & BGV mul. & 16 & 1024 & 27 & 32768 & 0.038 & 0.0 & 0.0 \\
SEAL & BGV mul. & 16 & 2048 & 54 & 32768 & 0.063 & 0.0 & 0.0 \\
SEAL & BGV mul. & 16 & 4096 & 72 & 32768 & 0.186 & 0.0 & 0.0 \\
SEAL & BGV mul. & 16 & 8192 & 174 & 32768 & 0.747 & 0.0 & 0.0 \\
\hline
DRAMatic & Combined & 509 & 1024 & 27 & 32768 & 0.171 & 0.025 & 0.032 \\
DRAMatic & Combined & 509 & 2048 & 54 & 32768 & 0.709 & 0.085 & 0.144 \\
DRAMatic & Combined & \textbf{509} & 4096 & 72 & 32768 & 2.398 & 0.217 & 0.399 \\
DRAMatic & Combined & \textbf{509} & 8192 & 174 & 32768 & 10.347 & 0.883 & 1.649 \\
SEAL & Combined & 16 & 1024 & 27 & 32768 & 0.184 & 0.0 & 0.0 \\
SEAL & Combined & 16 & 2048 & 54 & 32768 & 0.369 & 0.0 & 0.0 \\
SEAL & Combined & 16 & 4096 & 72 & 32768 & 1.291 & 0.0 & 0.0 \\
SEAL & Combined & 16 & 8192 & 174 & 32768 & 5.409 & 0.0 & 0.0 \\
\hline
\end{tabular}
}
\caption{Raw performance results of \autoref{fig:plot-dummy-mul-multiple} (using dummy multiplications).}
\label{tab:raw-results-10}
\end{table}

\end{document}